\documentclass[12pt,preprint,oneside,onecolumn,authoryear,longnamesfirst]{elsarticle} 

\usepackage{amsmath,amssymb,amsthm,amsfonts,amstext}
\usepackage{etoolbox}
\usepackage{lscape}
\usepackage{bm}
\usepackage{tabularx}
\usepackage{longtable}
\usepackage[flushleft]{threeparttable}
\appto\TPTnoteSettings{\linespread{1}\footnotesize} 
\makeatletter 
   \g@addto@macro\TPT@defaults{\linespread{1}\footnotesize} 
\makeatother
\usepackage{booktabs}
\usepackage{siunitx}
\usepackage{color}
\usepackage{graphicx}
\usepackage{caption}
\usepackage{subcaption}
\usepackage{mathrsfs}
\usepackage{setspace}
\usepackage{rotating}
\usepackage{multirow}

\usepackage{float}

\usepackage{tikz}
\usetikzlibrary{backgrounds,automata}
 
\usepackage{array}
\newcolumntype{$}{>{\global\let\currentrowstyle\empty}}
\newcolumntype{^}{>{\currentrowstyle}}
\newcolumntype{C}{%
  >{\rowstyle{}}c%
}
\newcolumntype{B}{%
  >{\currentrowstyle}S[detect-weight]%
}
\newcommand{\rowstyle}[1]{\gdef\currentrowstyle{#1}%
  #1\ignorespaces
}
\usepackage{thrmappendix}
\usepackage{alltt}
\usepackage{hypernat}
\usepackage{mathtools,dsfont}
\usepackage{hyperref}
\usepackage[capitalize,noabbrev]{cleveref}

\newtoggle{SUPPLEMENTAL}\toggletrue{SUPPLEMENTAL}
\togglefalse{SUPPLEMENTAL} 

\newtoggle{BLINDED}\toggletrue{BLINDED}
\togglefalse{BLINDED} 
\newcommand{\Blinded}[2]{\iftoggle{BLINDED}{#1}{#2}}

\usepackage[page]{appendix}

\theoremstyle{plain}
\newtheorem{assumption}{Assumption}

\newtheorem{theorem}{Theorem}
\newtheorem{lemma}[theorem]{Lemma}

\theoremstyle{definition}
\newcommand{\pconv}{\xrightarrow{p}}
\newcommand{\dconv}{\xrightarrow{d}}

\newcommand{\tr}{^{\top}}

\crefformat{footnote}{#2\footnotemark[#1]#3} 
\crefname{conjecture}{Conjecture}{Conjectures}
\crefname{section}{Section}{Sections}
\crefname{subsection}{Section}{Sections}
\crefname{subsubsection}{Section}{Sections}
\Crefname{conjecture}{Conjecture}{Conjectures}
\Crefname{section}{Section}{Sections}
\Crefname{subsection}{Section}{Sections}
\Crefname{subsubsection}{Section}{Sections}
\crefname{appendix}{Appendix}{Appendices}
\crefname{subappendix}{Appendix}{Appendices}
\crefname{subsubappendix}{Appendix}{Appendices}
\Crefname{appendix}{Appendix}{Appendices}
\Crefname{subappendix}{Appendix}{Appendices}
\Crefname{subsubappendix}{Appendix}{Appendices}
\crefname{equation}{}{}
\Crefname{equation}{Equation}{Equations}
\crefformat{enumi}{(#2#1#3)}
\crefrangeformat{enumi}{(#3#1#4)\crefrangeconjunction(#5#2#6)}
\crefmultiformat{enumi}{(#2#1#3)}{ and~(#2#1#3)}{, (#2#1#3)}{ and~(#2#1#3)}
\Crefformat{enumi}{Part (#2#1#3)}
\Crefrangeformat{enumi}{Parts (#3#1#4)\crefrangeconjunction(#5#2#6)}
\Crefmultiformat{enumi}{Parts (#2#1#3)}{ and~(#2#1#3)}{, (#2#1#3)}{ and~(#2#1#3)}
\crefname{assumption}{}{}
\Crefname{assumption}{Assumption}{Assumptions}
\newcommand{\crefrangeconjunction}{--}

\DeclareGraphicsExtensions{.pdf,.png}
\graphicspath{ {./Figures/} }

\newcommand{\citeposs}[1]{\citeauthor{#1}'s (\citeyear{#1})}

\newcommand{\matf}[1]{#1} 
\newcommand{\vecf}[1]{#1} 
\DeclareMathOperator{\Var}{Var}

\newcommand{\Varp}[1]{\Var\left(#1\right)}

\newcommand{\R}{{\mathbb R}}

\DeclareMathOperator{\E}{E}
\DeclareMathOperator{\Q}{Q_{\tau}}
\let\Pr\relax \DeclareMathOperator{\Pr}{P} 

\DeclareMathOperator*{\argmin}{arg\,min}

\DeclareMathOperator{\1}{\mathds{1}}
\newcommand{\Ind}[1]{\1\left\{#1\right\}}
\newcommand{\Normal}{\mathrm{N}}
\newcommand{\Normalp}[2]{\Normal\left(#1,#2\right)}

\newcommand{\UnifDist}{\textrm{Unif}}

\newcommand{\pD}[2]{\frac{\partial #1}{\partial #2}}

\providecommand{\absbig}[1]{\left\lvert#1\right\rvert}
\providecommand{\normbig}[1]{\left\lVert#1\right\rVert}
\providecommand{\abs}[1]{\lvert#1\rvert}
\providecommand{\norm}[1]{\lVert#1\rVert}
      
\let\originalleft\left
\let\originalright\right
\renewcommand{\left}{\mathopen{}\mathclose\bgroup\originalleft}
\renewcommand{\right}{\aftergroup\egroup\originalright}

\allowdisplaybreaks[3]

\usepackage[final]{listings}
\lstdefinestyle{inlineR}{language=R,frame=none,basicstyle=\ttfamily,keywordstyle=\ttfamily,stringstyle=\ttfamily,keepspaces=true,showspaces=false,showstringspaces=false,breaklines=true,upquote=true,print,columns=fullflexible}
\newcommand{\code}[1]{\lstinline@#1@}

\newtheorem{definition}[theorem]{Definition}

\def\N{\mathbb{N}}

\def\E{\textnormal{E}}

\def\Q{\textnormal{Q}_{\tau }}

\def\R{\mathbb{R}}

\def\X{\mathcal{X}}
\def\Y{\mathcal{Y}}

\def\pref{\succcurlyeq}

\def\kernel{K}

\def\R{\mathbb{R}}

\def\X{\mathcal{X}}
\def\Y{\mathcal{Y}}

\def\pref{\succcurlyeq}
\def\kernel{K}

\begin{document}

\raggedright
\parindent=1.5em
\doublespacing

\begin{frontmatter}
\title{A Smoothed GMM for Dynamic Quantile Preferences Estimation}
\tnotetext[disclosure]{Disclosure Statement: The authors declare that they have no competing interests.}
\tnotetext[data]{Data Availability Statement: The data and code that support the findings of this study are available in the supplemental materials.}

\author[add1]{\Blinded{[BLINDED]}{Xin Liu}}
\ead{\Blinded{[BLINDED]}{xin.liu1@wsu.edu}} 
\author[add2]{\Blinded{[BLINDED]}{Luciano de Castro}}
\ead{\Blinded{[BLINDED]}{decastro.luciano@gmail.com}}
\author[add3]{\Blinded{[BLINDED]}{Antonio F.\ Galvao}}
\ead{\Blinded{[BLINDED]}{agalvao@msu.edu}}
\address[add1]{\Blinded{[BLINDED]}{School of Economic Sciences, Washington State University, Pullman, WA 99164, United States.}}
\address[add2]{\Blinded{[BLINDED]}{Department of Economics, University of Iowa, Iowa City, IA 52242, United States.}}
\address[add3]{\Blinded{[BLINDED]}{Department of Economics, Michigan State University, East Lansing, MI 48824, United States.}}


\begin{abstract} 
\onehalfspacing{
This paper suggests methods for estimation of the $\tau$-quantile, $\tau\in(0,1)$, as a parameter along with the other finite-dimensional parameters identified by general conditional quantile restrictions. We employ a generalized method of moments framework allowing for non-linearities and dependent data, where moment functions are smoothed to aid both computation and tractability.
Consistency and asymptotic normality of the estimators are established under weak assumptions.
Simulations illustrate the finite-sample properties of the methods. An empirical application using a quantile intertemporal consumption model with multiple assets estimates the risk attitude, which is captured by $\tau$, together with the elasticity of intertemporal substitution.}
\end{abstract}


\end{frontmatter}

\thispagestyle{empty}
\doublespacing

\setcounter{page}{1}

\newpage

\section{Introduction}

A recent and expanding literature considers recursive economic models alternative to the standard expected utility (EU). In particular, \cite{deCastroGalvao19} suggested a dynamic quantile model.\footnote{The (static) quantile model was first introduced by \citet{Manski88}, and subsequently axiomatized by \citet{Chambers:09}, \citet{Rostek:10}, and \citet{deCastroGalvao22}.} In this framework, the economic agent chooses the alternative that leads to the highest $\tau$-quantile of the stream of future utilities for $\tau \in (0,1)$, where the parameter $\tau$ captures the risk attitude. The dynamic quantile preferences for intertemporal decisions are represented by an additively separable quantile model with standard discounting. 
The associated recursive equation is characterized by the sum of the current period utility function and the discounted value of the certainty equivalent, which is obtained from a quantile operator. This intertemporal model is tractable and simple to interpret, since the value function and Euler equation are transparent, and easy to calculate (analytically or numerically).
This framework allows the separation of the risk attitude from the intertemporal substitution, which is not possible with classic EU, while maintaining important features of the standard model, such as dynamic consistency and monotonicity. In the quantile model, the risk attitude is captured by $\tau$ and interpreted as downside risk.\footnote{There is an alternative existing literature examining downside risk, see, e.g., \citet{Rostek:10}, \citet{AngChenXing06}, and \citet{MenezesGeissTressler80}.}  Therefore, the model allows a separation of risk attitude (governed by $\tau$) and the elasticity of intertemporal substitution (EIS), 
which is exclusively determined by the utility function.\footnote{Quantile model is an alternative to EU models with useful advantages, such as, in the static case, robustness to monotonic transformations, allowing for a strong separation of beliefs and tastes. See \citet{deCastroGalvaoNunes25} for further details on dynamic quantile models.} 
Despite the growing interest in recursive utility models in economics, there has been a relatively small number of econometric work aimed at estimating the relevant parameters for recursive preferences. Some notable exceptions are, for example, \citet{HansenSingleton82}, \citet{EpsteinZin91}, \citet{HansenHeatonLeeRoussanov07}, and more recently \citet{ChenFavilukisLudvigson13}. Consequently, in empirical work, economic models are often calibrated with little econometric guidance regarding the value of preference parameters.


This paper contributes to the literature in multiple ways. 
First, we contribute to a growing literature on estimating recursive quantile preferences.
In such a model, the quantile $\tau$ is a structural parameter of interest, capturing the risk attitude. In the intertemporal consumption case, it would be important to identify and estimate both the risk attitude and EIS concurrently. 
Prior work (e.g., \citet{deCastroGalvaoKaplanLiu19}) is limited to estimating the EIS at different fixed risk attitudes indexed by $\tau$, instead of estimating the representative quantile.\footnote{\citet{YuZhang05} and \citet{BeraGalvaoMontesPark16} suggest parametric methods to estimate a quantile of interest based on the Asymmetric Laplace distribution.}
More recently, \citet{deCastroGalvaoOta24} provided methods to estimate the quantile parameter using a two-step procedure based on a conditional exogeneity assumption, --- but this approach does not allow for general endogeneity in the conditional quantile restriction. Our paper proposes methods for practical estimation of the quantile parameter $\tau$ together with other parameters of interest, adopting a smoothed GMM framework that has several advantages over existing approaches by allowing for endogenous regressors, non-linear models, weakly dependent data, and over-identified models.

Second, our framework allows for endogeneity. 
Differently from methods in \citet{deCastroGalvaoOta24}, which  rely on exogeneity of regressors and sample splitting, our methods allow for endogenous regressors. This is an important point because identification and estimation are based on the quantile Euler equation, which is only an equilibrium condition. It has become common in the literature to use such an equilibrium condition together with instrumental variables to derive orthogonality conditions that can be used to identify and estimate the parameters of interest (see, e.g., \citet{AttanasioLow04}). In this paper, when there are at least two choice variables in the recursive economic model, conditional on available instrumental variables, the parameters of interest satisfying both structural equilibrium moments are identifiable. Local identification follows directly from  \citet{ChenChernozhukovLeeNewey14}, since the equilibrium conditions can be written as usual moment conditions.

Given the identification of all the parameters of interest, moment conditions involving nonlinear structural quantile functions and instrumental variables are employed for estimation. 
The GMM optimization procedure minimizes the smooth quadratic objective function over the parameters of interest as well as the quantile parameter. 
We propose a two-step GMM procedure that is simple to be implemented in practice. Moreover, we provide conditions under which the proposed quantile smooth GMM estimator is $\sqrt{n}$-consistent, and establish its limiting distribution. Hence, the third main contribution is to improve on the computation and convergence rate of existing methods. 
The method in \citeposs{deCastroGalvaoOta24} computes the estimator using a grid search over quantiles, and the limiting results show a cubic-rate convergence rate of the estimator.
Instead, the proposed smoothed GMM estimator can directly estimate the quantile parameter solving a minimization problem, attaining a parametric rate of convergence, i.e., it is $\sqrt{n}$-consistent instead of cubic rate.

Fourth, in a broader econometric view, this paper extends the existing literature on instrumental variables quantile regression (IVQR).
In prior work, 
\citet{ChernozhukovHansen05, ChernozhukovHansen06, ChernozhukovHansen08} provide results on identification, estimation, and inference for a quantile model that allows for endogenous regressors;
\citet{KaplanSun17} show that smoothing improves computation and estimation efficiency in IVQR model. 
\citet{deCastroGalvaoKaplanLiu19} further extends to a non-linear QR model and allow for weakly dependent data in a GMM framework. See \citet{ChernozhukovHansenWuthrich17} for an overview of IVQR.
All of these prior work focus on estimating the parameter of interest at fixed quantile level. 
Instead, our framework treat the quantile level as a structural parameter, which is a single quantile that satisfies the equilibrium condition of multiple choice variables. 
This structural quantile parameter can be identified and estimated when there are at least two choice variable moments.
We extend \citeposs{deCastroGalvaoKaplanLiu19} smoothed GMM framework to estimate the structural quantile parameter. As a result and mentioned above, our framework has the advantages of allowing for endogenous regressors, weakly dependent data, and non-linear and over-identified models.

Built on the general structural dynamic quantile model, we provide an empirical application to an intertemporal consumption model with multiple assets to illustrate the estimation. Using US data, it jointly estimates the EIS and the risk attitude, which is captured by the quantile. 
We find very sensible estimates. Results show empirical evidence of a slight risk aversion quantile parameter, and EIS slightly smaller than unit. These results are in line with the literature, where risk aversion is documented (see, e.g., \citet{ElminejadHavranekIrsova22}, and \citet{deCastroGalvaoNoussairQiao21}), and EIS relatively close to one (see, e.g., \citet{Havranek15}, \citet{Thimme17}, and \citet{Bestetal20}) is also reported.

Finally, we provide Monte Carlo simulations evaluating the finite sample properties of the proposed methods. Numerical results provide evidence that the estimator for the parameters of interest is approximately unbiased. In particular, the GMM method is able to estimate the risk attitude parameter well in finite samples.

The remaining of the paper is as follows. Section \ref{sec:Econ} introduces the structural economic quantile model, and Section \ref{sec:moments} provides the relevant moment conditions for estimation. In Section \ref{sec:est}, we present the GMM estimators. Section \ref{sec:asy} provides the limiting statistical properties of the estimators. An empirical application is given in Section \ref{sec:Application}, and a Monte Carlo simulation in Section \ref{sec:MC}. Finally, Section \ref{sec:Conclusion} concludes. All proofs are collected in the Appendix.

Notation. 
Random variables and vectors are uppercase ($Y$, $X$, etc.), while non-random values are lowercase ($y$, $x$); 
for vector/matrix multiplication, all vectors are treated as column vectors. 
Also, $\Ind{\cdot}$ is the indicator function, 
$\E(\cdot)$ expectation, 
$\Q(\cdot)$ the $\tau$-quantile, 
$\Pr(\cdot)$ probability, 
and $\Normalp{\mu}{\sigma^2}$ the normal distribution. 
For vectors, $\normbig{\cdot}$ is the Euclidean norm. 
Acronyms used include those for 
cumulative density function (CDF),
central limit theorem (CLT), 
elasticity of intertemporal substitution (EIS), 
generalized method of moments (GMM), 
probability density function (PDF), 
root mean squared error (RMSE),
structural quantile functions (SQF), and
uniform law of large numbers (ULLN). 

\section{The Economic Model}
\label{sec:Econ} 

This section reviews the general dynamic quantile model for a representative agent studied by \citet{deCastroGalvaoNunes25}. The main objective is to obtain a general quantile Euler equation that serves as basis for identification and estimation of the $\tau$-quantile parameter of interest, along with the other parameters characterizing the preferences. Most of what follows summarizes the model discussed in that paper.


Let $\X$ denote the state space, $\Y$ be the set of possible actions the decision-maker (DM) may take, and $\mathcal{E}$, the range of the shocks (random variables) in the model. We require these sets to be metric spaces.
Let $x_{t} \in \X$ denote the state in period $t$, and $e_{t} \in \mathcal{E}$ the shock after the end of period $t-1$, both of which are known by the DM at the beginning of period $t$. 
In each period $t$, the DM chooses a feasible action $y_t$ from a constraint subset $\Gamma (x_{t},e_{t}) \subset \Y$, that is, the feasibility constraint set.
The choice $y_{t}$ and the shock $e_{t+1}$ affect next period state $x_{t+1}$ through  a law of motion function $\phi$ from $\X \times \Y \times \mathcal{E}  $ to $\X$, that is, 
\begin{equation}\label{eq:law of motion}
x_{t+1} = \phi(x_t, y_t,  e_{t+1}).
\end{equation}
In many models, the DM chooses directly the next state, in which case $\phi(x_t,y_t,e_{t+1})=y_t$.  
 

The random shocks will follow a time-invariant (stationary) Markov process. We particularize the setting in \citet{deCastroGalvaoNunes25} and assume that the set of  random shocks $\mathcal{E}$ is  an interval. 
Stationary Markov processes are modeled by a Markov kernel $\kernel: \mathcal{E} \times \Sigma \to [0,1]$, where $\Sigma$ is the Borel $\sigma$-algebra of the metric space $\mathcal{E}$.\footnote{Recall that a mapping $\kernel:\mathcal{E} \times \Sigma \to [0,1]$ is a Markov kernel if 
for each $e \in \mathcal{E}$, the set function $\kernel(e,\cdot) : \Sigma \to [0,1]$ is a probability measure and, for each $S \in \Sigma$, the mapping $\kernel(\cdot, S) \to [0,1]$ is $\Sigma$-measurable. See \citet[Definition  19.11, p. 630]{Aliprantis_Border_2006}.} 
This means that  the probability that $A'\in B \subset \mathcal{E}$ given $A=e$ is $\Pr \left( A'\in B | A=e \right) = \kernel(e, B)$. 
The expectation of  a function $h: \mathcal{E} \to \R$ is 
$    \E\left[h(w) \big| \, e \right] = \int_{\mathcal{E}} h(e') k(e, de')$.




\subsection{The Recursive Problem}

Given the current state   $x_t$ and current shock $e_{t}$,  $\Gamma (x_t,e_{t})$ denotes the  set of possible choices $y_{t}$. Given $x_t, e_{t}$ and $y_{t} \in \Gamma(x_t, e_{t})$, $u\left(x_{t}, y_{t}, e_{t} \right)$ denotes the instantaneous utility obtained in period $t$. Let $e_{t} \subset \mathcal{F}_{t}$ and $\{\mathcal{F}_{t}\}_{t=0}^{\infty}$ denote the sequence of increasing conditioning information sets available to a representative economic agent. 
In our model, the uncertainty with respect to the future realizations of $e_{t}$ are evaluated by a quantile. 
In the dynamic quantile model, the intertemporal choices can be represented by the maximization of a value function $v : \X \times \mathcal{E} \to \R$ that satisfies the recursive equation: 
\begin{equation}\label{eq:FE}
v(x,e)= 
\sup_{
y \in \Gamma (x,e) }
\; \; 
\Bigl\{  u\left(x, y , e \right) \; + \;  \delta \Q  [  \; v \left( \phi(x,y,e')
, \, e' \right) \; | \; \mathcal{F}] \Bigr\},
\end{equation}
where $e'$ indicates the next period shock, $\delta$ is the discount factor parameter, $\tau$ is a parameter capturing the risk attitude. Additional details on the risk attitude for quantile preference models are provided in Appendix \ref{app:preferences}. 

In the dynamic quantile model in equation \eqref{eq:FE} the DM maximizes the current period utility plus the $\tau$-quantile of the uncertain discounted stream of future utilities. Intuitively, the $\tau$-quantile of a random variable is the value that is exceeded with probability $(1-\tau)$, for $\tau\in(0,1)$, such that the quantile captures the downside risk attitude. 
This problem modifies the usual dynamic programming problem by substituting the expectation operator $\E[\cdot]$  by the quantile $\Q [\cdot]$. 
See \citet{deCastroGalvao19} for a construction of this recursive model from dated preferences.

\subsection{Assumptions}\label{sec:assumptions}

Now we state assumptions on the economic model that allows for deriving the Euler equation. 
For discussions and justifications of these conditions, see \citet{deCastroGalvaoNunes25}.

\begin{assumption}\label{ass:basic}
The following holds:
\begin{enumerate}[(1)]
    \item $\mathcal{E} \subseteq \R$ is an interval;
    \item The  transition function  
$\kernel:\mathcal{E} \times \Sigma \to [0,1]$ satisfies  the following:
\begin{enumerate}[(i)]
\item for each $e \in \mathcal{E}$ and $ \eta \in (0,1)$, there exists  compact $\mathcal{E}' \subset \mathcal{E}$ such that
$
    \kernel(e,\mathcal{E}') > 1 - \eta;
$
\item for each compact $A \subset \mathcal{E}$, the function
$
    e \in \mathcal{E} \mapsto \kernel(e,A) \in [0,1]  
$ 
is continuous;

\item for each $A \in \Sigma$  open and nonempty,
$\kernel(e,A) > 0 \text{ for all } e \in \mathcal{E}$;

\item 
for any weakly increasing function $h: \mathcal{E} \to \R$ and $e,e' \in \mathcal{E}$ such that     
 $e \leqslant e'$,
 $\E \left[h(w) \big| \, e\right] \leqslant  \E \left[h(w) \big| \, e'\right]$. 
\end{enumerate}

\item The discount rate $\delta$ lies between $0$ and $1$, that is, $\delta \in (0,1)$;

\item $\X \subset \R^p$ is convex; and  $\Y \subset \R^m$ is convex;

\item The correspondence  $\Gamma: \X \times \mathcal{E} \rightrightarrows \Y $ is continuous, with nonempty, compact values.
Moreover, for every $e\in \mathcal{E}$ and $x \leqslant x'$, $\Gamma (x, e) \subseteq \Gamma (x',e)$ and $\Gamma$ is convex; 

\item The utility function  $u: \X \times \Y \times \mathcal{E} \to \R$ is continuous, bounded, and  $C^1$ in $x$.
Moreover $u$ is non-decreasing in $x$,  concave in $(x,y)$ , and non-decreasing in $e$; 

 \item $\phi: \X \times \Y \times \mathcal{E} \to \X$ is differentiable  and   does not depend on $x$. Moreover,  $\phi$ is  concave in $(x,y)$ and non-decreasing in $e$. 

\end{enumerate}
\end{assumption}

\subsection{Euler Equation}\label{sec:Euler equation}

Under these  assumptions, \citet{deCastroGalvaoNunes25} establish the following Euler equation result for the choice variable $y_{j}$, $j=1,..,m$:

\begin{theorem}[Euler Equation; \citet{deCastroGalvaoNunes25}]\label{th:Euler equations}
Let Assumption \ref{ass:basic} hold. 
Let $(x_t,y_t,e_t)_{t\in \N}$ be a sequence of states, optimal decisions and shocks, such that $(x_t,y_{t})$ are interior for all $t$.
If $e_{t} \mapsto \frac{\partial u}{\partial x}\left(x_{t} , y_{t} , e_{t} \right) \cdot \frac{\partial \phi}{\partial y_{j}}\left(y_{t-1} , e_{t} \right)$ is strictly increasing, then $\forall t\in \N$ and $j=1,...,m$:
\begin{equation}\label{eq:Euler suming dimensions}
    \Q  \left[ \frac{\partial u}{\partial y_{j}} \left(x_{t} , y_{t} , e_{t} \right) + \delta
  \sum_{l=1}^p \frac{ \partial u } {\partial x_{l} }  \left( x_{t+1} ,y_{t+1} ,  e_{t+1}\right)  \frac{ \partial \phi_{l} } {\partial y_{j}}   
  \left(y_{t} ,  e_{t+1}\right) \Big| \, \mathcal{F}_{t}
  \right]
   =0, 
\end{equation}
where $\phi_{l}$ stands for the $l$-th component of $\phi$.  
\end{theorem}

\medskip

In equation \eqref{eq:Euler suming dimensions}, $\frac{\partial u}{\partial {y_{j}}}$ represents the derivative of $u$ with respect to the $j$-th coordinate of its second variable ($y$) (that is, an unidimensional value) and $\frac{\partial u}{\partial x}$ represents the derivative of $u$ with respect to  its first variable ($x$) (that is, a $p$-dimensional vector). Since $\phi$ takes value on $\X \subset \R^{p}$, $\frac{\partial \phi}{\partial{y_{j}}}$ stands for the $p$-dimensional derivative vector of $\phi$ with respect to the $j$-th coordinate of $y$. The conditional quantile function in \eqref{eq:Euler suming dimensions} is valid for multiple choice variables $j=1,...,m$. When $m\geq 2$, one is able to use the corresponding moment conditions to recover the $\tau$ quantile parameter, in addition to the other parameters characterizing the preferences. We discuss these details in the next section, and provide an empirical example using intertemporal consumption with multiple assets in Section \ref{sec:Application} below.

Theorem \ref{th:Euler equations} provides a general version of the Euler equation, that is the optimality conditions for the quantile dynamic programming problem. The Euler equation in \eqref{eq:Euler suming dimensions} is displayed as an implicit function, nevertheless for any particular application, and given utility function, one is able to solve it explicitly as a conditional quantile function.

When $\phi(x, y,e) = y$ and $\X \equiv \Y$, as in the model where DM chooses an action that is in the set of next period state, \eqref{eq:Euler suming dimensions} simplifies to
\begin{equation}\label{eqn:special-Euler}
      \Q \left[ \frac{\partial u}{\partial y_{j}} \left(x_{t} , y_{t} , e_{t} \right) + \delta\frac{\partial u}{\partial x_{j}} \left( x_{t+1} ,y_{t+1} ,  e_{t+1}\right) \Big| \, \mathcal{F}_{t}\right]
   =0,
\end{equation}
for $j=1,...,m$.

\section{Moment conditions representation}\label{sec:moments}

The results in Theorem \ref{th:Euler equations}, and more generally results in \citet{deCastroGalvaoNunes25}, show that the equilibrium condition for the quantile economic model can be written as an Euler equation, which in turn can be described by a non-linear conditional quantile function.
These quantile functions can be represented in terms of conditional moments as discussed below. Such a representation suggests the use of a GMM method as a strategy to recover the parameters of interest.

An important feature of the Euler equation in \eqref{eq:Euler suming dimensions} is that it is only an equilibrium condition. In the literature, it has become common to use such an equilibrium condition together with instrumental variables to derive orthogonality conditions that can be used to identify and estimate the parameters of interest (see, e.g., \citet{AttanasioLow04}). Nevertheless, this quantile Euler equation alone is not able to allow for identification and estimation of the $\tau$ quantile, i.e., the risk attitude parameter. 
Thus, at least two observable choice variables together with the quantile Euler equation allow for identification of the parameter $\tau$.

Now, we consider a general nonlinear model representing \eqref{eq:Euler suming dimensions}.
For simplicity of exposition, from now on, we restrict to two choices $j=1,2$. But results generalize to a given number of moment conditions $m$, as long as $m\ge 2$.

From the quantile Euler equation in \eqref{eq:Euler suming dimensions}, we have the following nonlinear conditional quantiles for both choices $j=1,2$,
\begin{equation}\label{eq:qrmodel0}
\textnormal{Q}_{\tau_{0}}[ \Lambda(Y_{ji}, X_{ji},\beta_0)  \mid Z_{ji}]  = 0, 
\end{equation}
where 
$Y_{ji}\in\mathcal{Y}\subseteq\R^{d_Y}$ is the endogenous variable vector, 
specifically, $Y_{ji}=(Y_{1,ji}, Y_{-1,ji})$ consists of both the outcome $Y_1$ and the endogenous regressors $Y_{-1}=(Y_2,\ldots,Y_{d_Y})$,
$Z_{ji}\in\mathcal{Z}\subseteq\R^{d_Z}$ is the full instrument vector that contains $X_{ji}\in\mathcal{X}\subseteq\R^{d_X}$ as a subset, and $Z \subset \mathcal{F}$. 
The parameters of interest are the finite-dimensional $\beta_0 \in \mathcal{B} \subseteq \R^{d_\beta}$ 
and the quantile index $\tau_0 \in \mathcal{T} \subset (0,1)$. 
The $\Lambda(\cdot)$ is the known structural function. With instruments $Z$, the structural function can be identified and estimated.
We highlight that this structure is an extension of the literature, as it can address endogeneity in the structural function, whereas prior work, such as \citet{deCastroGalvaoOta24}, relies on exogenous regressors.
We consider a general econometric framework that not only works for iid data over $i$, but also 
allows for weakly dependence across $i$, in which case the $i$ can represent the $t$ in \cref{eq:Euler suming dimensions}.
Note that we use the term ``conditional quantile function'' in \cref{eq:qrmodel0} to represent the conditional quantile of the structural function, given the instruments. This is different from the conditional quantile function in the econometric sense, where there is no endogeneity.

The zero conditional quantile function in equation \eqref{eq:qrmodel0} can be written as the following conditional expectation
\begin{equation}\label{eq:condE}
\E \left[ \Ind{\Lambda\bigl(\vecf{Y}_{ji},\vecf{X}_{ji},\vecf{\beta}_0 \bigr) \le 0} - \tau_0 \mid \vecf{Z}_{ji} \right]=0,\ j=1,2,
\end{equation}
where $\Ind{\cdot}$ is the indicator function. By the law of iterated expectation, \cref{eq:condE} implies the following unconditional expectation
\begin{equation}\label{eq:qrgmmmodel}
\E \left[ \vecf{Z}_{ji} (\Ind{\Lambda\bigl(\vecf{Y}_{ji},\vecf{X}_{ji},\vecf{\beta}_0 \bigr) \leq 0} - \tau_0 ) \right]=0,\ j=1,2, 
\end{equation}
which is more explicitly written as
\begin{align}\label{eqn:moments}
\E
\begin{bmatrix}
\vecf{Z}_{1i} \left[ \Ind{\Lambda\bigl(\vecf{Y}_{1i},\vecf{X}_{1i},\vecf{\beta}_0 \bigr) \le 0} - \tau_0 \right]\\
\vecf{Z}_{2i} \left[ \Ind{\Lambda\bigl(\vecf{Y}_{2i},\vecf{X}_{2i},\vecf{\beta}_0 \bigr) \leq 0} - \tau_0 \right]
\end{bmatrix}=
\begin{bmatrix}
    0 \\
    0
\end{bmatrix}. 
\end{align}
Note that the parameters $(\beta_0, \tau_0)$ solve the moment conditions \cref{eq:qrgmmmodel} or \cref{eqn:moments} exactly. 


Let $M(\beta,\tau)=\E
\begin{bmatrix}
\vecf{Z}_{1i} \left[ \Ind{\Lambda\bigl(\vecf{Y}_{1i},\vecf{X}_{1i},\vecf{\beta} \bigr) \le 0} - \tau \right]\\
\vecf{Z}_{2i} \left[ \Ind{\Lambda\bigl(\vecf{Y}_{2i},\vecf{X}_{2i},\vecf{\beta} \bigr) \leq 0} - \tau \right]
\end{bmatrix}$. 
From the moment condition in equation \eqref{eqn:moments}, and assuming that $M(\beta,\tau)$ is differentiable at $(\beta_0,\tau_0)$, and the derivative of $M(\beta,\tau)$ at $(\beta_0,\tau_0)$ has full rank, then the parameters of interest, $(\beta_0,\tau_0)$, are locally identified. This result is a direct application of \citet{ChenChernozhukovLeeNewey14}.


\section{The smoothed {GMM} estimators}
\label{sec:est}

This section introduces the econometric model we suggest to estimate the parameters of interest. For simplicity, we restrict our attention to a model with two choice variables, but the extension to multiple choices is straightforward.

Let the population map $M \colon \mathcal{B} \times \mathcal{T} \mapsto \R^{2d_Z}$ be 
\begin{align}\label{eqn:def-M}
M(\beta,\tau) 
&\equiv \E\left[ g^u_i(\beta,\tau) \right] 
,\\ \label{eqn:def-gu}
g^u_i(\beta,\tau)
&\equiv g^u(Y_i,X_i,Z_i,\beta,\tau) \notag\\
& \equiv 
\begin{bmatrix}
\vecf{Z}_{1i} (\Ind{\Lambda\bigl(\vecf{Y}_{1i},\vecf{X}_{1i},\vecf{\beta} \bigr) \leq 0} - \tau )\\
\vecf{Z}_{2i} (\Ind{\Lambda\bigl(\vecf{Y}_{2i},\vecf{X}_{2i},\vecf{\beta} \bigr) \leq 0} - \tau )
\end{bmatrix}, 
\end{align}
where superscript ``$u$'' denotes ``unsmoothed,''
$Y_i=(Y_{1i}',Y_{2i}')'$, $X_i=(X_{1i}',X_{2i}')'$, and $Z_i=(Z_{1i}',Z_{2i}')'$. 
We have $2d_Z$ moments to solve for $d_\beta+1$ parameters. 
We assume $d_Z\ge d_\beta$, because for a single choice variable we assume the number of instruments ($d_Z$) should be at least equal to or greater than the number of parameters $(d_\beta)$ in order to identify the parameter $\beta$ in the conventional fixed-$\tau$ quantile regression model. 
This condition for a single choice variable is sufficient for the (over-)identification ($2d_Z \ge d_\beta + 1 $) when two choice variables are available.
As derived in \eqref{eqn:moments}, the population moment condition is 
\begin{equation}\label{eqn:def-beta0}
0 = M(\beta_0,\tau_0) . 
\end{equation}

With the unsmoothed sample moments $\hat{M}_n^u\left(\beta,\tau\right)\equiv 1/n \sum_{i=1}^{n} g^u_i(\beta,\tau) $, 
which has a discontinuous indicator function $\Ind{\cdot}$ inside,
it is computationally expensive to search for the minimizing point of the GMM objective function, especially in the case of multiple endogenous regressors. 
Following \citet{deCastroGalvaoKaplanLiu19}, we use the smoothed GMM method, replacing the indicator function with a smoothed function $\tilde{I}(\cdot)$.

Specifically, we use the smoothed sample moments (no ``$u$'' superscript),
\begin{align}
\begin{split}\label{eqn:def-M-hat}
g_{ni}(\beta,\tau)
&\equiv g_n(Y_i,X_i,Z_i,\beta,\tau)\\
& \equiv 
\begin{bmatrix}
Z_{1i} \bigl[ \tilde{I}\bigl(-\Lambda(Y_{1i},X_{1i},\beta) / h_n \bigr) - \tau \bigr]\\
Z_{2i} \bigl[ \tilde{I}\bigl(-\Lambda(Y_{2i},X_{2i},\beta) / h_n \bigr) - \tau \bigr]
\end{bmatrix} \\
\hat{M}_n\left(\beta,\tau\right)
&\equiv \frac{1}{n}\sum_{i=1}^{n} g_{ni}(\beta,\tau) , 
\end{split}
\end{align}
where 
$h_n$ is a bandwidth (sequence).
We use the same smoothed function $\tilde{I}(\cdot)$\footnote{$\tilde{I}(u)=\Ind{-1\leq u\leq 1}[0.5+\frac{105}{64}(u-\frac{5}{3}u^3+\frac{7}{5}u^5-\frac{3}{7}u^7)]+\Ind{u>1}$.} as in \citet{Horowitz98}, \citet{Whang06}, \citet{deCastroGalvaoKaplanLiu19}, and \citet{KaplanSun17}, 
which has shown high-order improvements in a linear iid setting. 
Using the same notations as in \citet{deCastroGalvaoKaplanLiu19},
the double subscript on $g_{ni}$ is a reminder of the triangular array setup since $g_{ni}$ depends on $h_n$ in addition to $(Y_i,X_i,Z_i)$. Next, we describe the estimator.

\subsection{One-step estimator}
\label{sec:est-1s}

We briefly introduce a one-step estimator that serves as an intermediate step in the two-step GMM estimator (introduced in \cref{sec:est-GMM}) that we focus on and establish the large-sample theory in \cref{sec:asy}.

This one-step estimator is parallel to \citeposs{deCastroGalvaoKaplanLiu19}, and it originally follows from \Citet[eq. (3.11), p.\ 2151]{NeweyMcFadden94}.
That is,
\begin{equation}\label{eqn:def-est-1s}
(\hat{\beta}_{\mathrm{1s}}\tr,\hat{\tau}_{\mathrm{1s}})\tr
= (\bar{\beta}\tr,\bar{\tau})\tr 
  -(\bar{\matf{G}}\tr \bar{\matf{\Sigma}}^{-1}\bar{\matf{G}})^{-1}
  \bar{\matf{G}}\tr \bar{\matf{\Sigma}}^{-1}
  \sum_{i=1}^{n}g_{ni}(\bar{\beta},\bar{\tau})/n  
,
\end{equation}
where $(\bar{\beta},\bar{\tau})$ is some initial consistent (but not efficient) estimator. The $\bar{\matf{G}}=\frac{1}{n} \sum_{i=1}^{n} \nabla_{(\beta\tr,\tau)} g_{ni}(\bar{\beta},\bar{\tau})$ , 
where $\nabla_{(\beta\tr,\tau)}$ denotes the partial derivative with respect to $(\beta\tr,\tau)$.
The $\bar{\matf{\Sigma}}$ is a consistent estimator of 
$\matf{\Sigma} 
 = \E[ g_{ni}( \beta_0, \tau_0 ) g_{ni}( \beta_0, \tau_0 )\tr ]$. 
With iid data, let $\bar{\matf{\Sigma}}=\frac{1}{n}\sum_{i=1}^{n} g_{ni}(\bar{\beta},\bar{\tau}) g_{ni}(\bar{\beta},\bar{\tau})\tr $.
With dependent data, we use a long-run variance estimator $\bar{\matf{\Sigma}}$ as in \citet{NeweyWest87} and \citet{Andrews91}.
\Citet[Thm.\ 3.5, p.\ 2151]{NeweyMcFadden94} show that 
this one-step estimator achieves asymptotic efficiency, although they assume the moment function $g(\cdot)$ is fixed (not depending on $n$).

\subsection{Two-step {GMM} estimator}
\label{sec:est-GMM}

We first define a generic smoothed GMM estimator that minimizes a quadratic form of the smoothed sample moments with a generic weighting matrix,
\begin{align}\label{eqn:def-est-GMM}
(\hat{\beta}_{\mathrm{GMM}},\hat{\tau}_{\mathrm{GMM}}) 
    & =\argmin_{ (\beta, \tau)\in \mathcal{B}\times \mathcal{T} } 
    \left[ \sum_{i=1}^{n} g_{ni}(\beta, \tau) \right]\tr \hat{\matf{W} } \left[ \sum_{i=1}^{n} g_{ni}(\beta, \tau) \right]  \notag \\
     & =\argmin_{ (\beta, \tau)\in \mathcal{B}\times \mathcal{T} } 
      \hat{M}_n(\beta,\tau)\tr  \hat{\matf{W} }   \hat{M}_n(\beta,\tau),
\end{align}
where $\hat{\matf{W}}$ is a generic symmetric, positive definite weighting matrix. Notice that the optimization is taken over all parameters, including the quantile $\tau$ as a parameter to be estimated.

When $\hat{\matf{W}}$ is the optimal weighting matrix, 
(for example, it is an estimator of the inverse long-run variance of the sample moments $\hat{\matf{W}}^* = \bar{\matf{\Sigma}}^{-1} \pconv \matf{\Sigma}^{-1}$,)
we define the efficient two-step GMM estimator as
\begin{equation}\label{eqn:def-est-2s}
(\hat{\beta}_{\mathrm{2s}},\hat{\tau}_{\mathrm{2s}})
= \argmin_{
(\beta, \tau)\in \mathcal{B}\times \mathcal{T}
} 
    \hat{M}_n(\beta,\tau)\tr  \bar{\matf{\Sigma}}^{-1}
    \hat{M}_n(\beta,\tau)
,
\end{equation}
where we use the one-step estimator in \cref{sec:est-1s} to compute the efficient weighting matrix $\bar{\matf{\Sigma}}^{-1}$.

Although smoothed, the GMM objective function may still be non-convex, which makes the computation difficult. 
Following \citet{deCastroGalvaoKaplanLiu19}, we use the simulated annealing algorithm from the GenSA package in R to search for the global minimum. 
This algorithm has the advantage of escaping local minima and exploring a wider search space, which is especially useful for non-convex objective functions.
However, this algorithm relies on an initial value reasonably close to the solution.
Thankfully, we use the one-step estimator in \cref{sec:est-1s} as the initial value.

\section{Large sample properties}
\label{sec:asy}

This section establishes the limiting properties of the proposed estimator. We first introduce the assumptions and then present the consistency and asymptotic normality of the smoothed GMM estimator.

\subsection{Assumptions}
\label{sec:asy-assumptions}

\begin{assumption}\label{a:XYZ}
For each observation $i$ among $n$ in the sample, for two choices $j=1,2$, endogenous vector $Y_{ji}\in\mathcal{Y}\subseteq\R^{d_Y}$ and instrument vector $Z_{ji}\in \mathcal{Z}\subseteq\R^{d_Z}$; a subset of $Z_{ji}$ is $X_{ji}\in\mathcal{X}\subseteq\R^{d_X}$, with $d_X\le d_Z$. 
The sequence $\{Y_{ji},Z_{ji}\}$ is strictly stationary and weakly dependent. 
\end{assumption}
\begin{assumption}\label{a:Lambda}
The function $\Lambda \colon \mathcal{Y}\times\mathcal{X}\times\mathcal{B} \mapsto \R$ is known and has (at least) one continuous derivative in its $\mathcal{B}$ argument for all $y\in\mathcal{Y}$ and $x\in\mathcal{X}$. 
\end{assumption}
\begin{assumption}\label{a:B}
The parameter space $\mathcal{B}\times \mathcal{T} \subset\R^{d_\beta}\times  (0,1)$ is compact; $d_\beta \le d_Z$ (which implies $d_\beta+1 \le 2d_Z$).
The population parameter $(\beta_0,\tau_0)$ is in the interior of $\mathcal{B} \times \mathcal{T}$ and uniquely satisfies the moment condition 
\begin{equation}\label{eqn:ID}
0 
= \E\begin{bmatrix}
\vecf{Z}_{1i} (\Ind{\Lambda\bigl(\vecf{Y}_{1i},\vecf{X}_{1i},\vecf{\beta}_0 \bigr) \leq 0} - \tau_0 )\\
\vecf{Z}_{2i} (\Ind{\Lambda\bigl(\vecf{Y}_{2i},\vecf{X}_{2i},\vecf{\beta}_0 \bigr) \leq 0} - \tau_0 )
\end{bmatrix} . 
\end{equation}
\end{assumption}
\begin{assumption}\label{a:Z}
The matrix $\E\left(Z_i Z_i\tr \right)$ is positive definite (and finite), where $Z_i=(Z_{1i}\tr, Z_{2i}\tr)\tr$.
\end{assumption}
\begin{assumption}\label{a:Itilde}
The function $\tilde{I}(\cdot)$ satisfies $\tilde{I}(u)=0$ for $u\le-1$, $\tilde{I}(u)=1$ for $u\ge1$, and $-1\le\tilde{I}(u)\le2$ for $-1<u<1$. 
The derivative $\tilde{I}'(\cdot)$ is a symmetric, bounded kernel function of order $r\ge2$, so $\int_{-1}^{1}\tilde{I}'(u)\,du=1$, $\int_{-1}^{1}u^k\tilde{I}'(u)\,du=0$ for $k=1,\ldots,r-1$, and $\int_{-1}^{1}\lvert u^r\tilde{I}'(u) \rvert \,du<\infty$ but $\int_{-1}^{1}u^r\tilde{I}'(u)\,du\ne0$, and $\int_{-1}^{1} \lvert u^{r+1} \tilde{I}'(u) \rvert \,du<\infty$. 
\end{assumption}
\begin{assumption}\label{a:h}
The bandwidth sequence $h_n$ satisfies $h_n=o(n^{-1/(2r)})$. 
\end{assumption}
\begin{assumption}\label{a:Y}
Given any $\beta\in\mathcal{B}$ and almost all $Z_{ji}=z$ (i.e.,\ up to a set of zero probability), 
the conditional distribution of $\Lambda(Y_{ji},X_{ji},\beta)$ given $Z_{ji}=z$ is continuous in a neighborhood of zero, for $j=1,2$. 
\end{assumption}
\begin{assumption}\label{a:ULLN}
Using the definition in \cref{eqn:def-M-hat}, 
\begin{equation}\label{eqn:smooth-Mn-ULLN}
\sup_{(\beta, \tau)\in \mathcal{B}\times \mathcal{T}
}
\norm{ \hat{M}_n(\beta,\tau) - \E[ \hat{M}_n(\beta,\tau) ] }
= o_p(1) . 
\end{equation}
\end{assumption}
\begin{assumption}\label{a:U}
Let $\Lambda_{ji}\equiv \Lambda\bigl(Y_{ji},X_{ji},\beta\bigr)$ and $(D_{ji}\tr,1)\tr \equiv \nabla_{
(\beta\tr,\tau)\tr
} (\Lambda\bigl(Y_{ji},X_{ji},\beta\bigr)-\tau)$, 
using the notation
\begin{equation}\label{eqn:def-nabla-b-Lambda}
\nabla_{(\beta\tr,\tau)\tr} \left( \Lambda\left(y,x,\beta\right) -\tau \right)
\equiv \pD{}{(\beta\tr,\tau)\tr} \left( \Lambda\left(y,x,\beta\right) -\tau \right) \Bigr\rvert_{\beta=\beta_0, \tau=\tau_0} , 
\end{equation}
for the $(d_\beta+1)\times1$ partial derivative vector, $j=1,2$. 
Let $f_{\Lambda_j|Z_j}(\cdot\mid z)$ denote the conditional PDF of $\Lambda_{ji}$ given $Z_{ji}=z$, and let $f_{\Lambda_j|Z_j,D_j}(\cdot\mid z,d)$ denote the conditional PDF of $\Lambda_{ji}$ given $Z_{ji}=z$ and $D_{ji}=d$, for $j=1,2$. 
(i) For almost all $z$ and $d$, $f_{\Lambda_j|Z_j}(\cdot\mid z)$ and $f_{\Lambda_j|Z_j,D_j}(\cdot\mid z,d)$ are at least $r$ times continuously differentiable in a neighborhood of zero, where the value of $r$ is from \cref{a:Itilde}. 
For almost all $z\in\mathcal{Z}$ and $u$ in a neighborhood of zero, there exists a dominating function $C(\cdot)$ such that $\abs{f_{\Lambda_j|Z_j}^{(r)}(u\mid z)}\le C(z)$ and $\E\bigl[C(Z_j)\abs{Z_j}\bigr] < \infty$, for $j=1,2$. 
(ii) The matrix
\begin{align}\label{eqn:def-G}
\matf{G} & \equiv \frac{\partial}{\partial (\beta\tr,\tau)} \E\begin{bmatrix}
\vecf{Z}_{1i} (\Ind{\Lambda\bigl(\vecf{Y}_{ji},\vecf{X}_{1i},\vecf{\beta} \bigr) \leq 0} - \tau )\\
\vecf{Z}_{2i} (\Ind{\Lambda\bigl(\vecf{Y}_{ji},\vecf{X}_{2i},\vecf{\beta} \bigr) \leq 0} - \tau )
\end{bmatrix}
\Bigr\rvert_{\beta=\beta_0,\tau=\tau_0}  \\
 & = 
 \begin{bmatrix}
-\E\left[ Z_{1i} D_{1i}\tr   f_{\Lambda_1|Z_1,D_1}(0\mid Z_{1i},D_{1i})  \right] & - \E\left[Z_{1i} \right]\\
-\E\left[ Z_{2i} D_{2i}\tr   f_{\Lambda_2|Z_2,D_2}(0\mid Z_{2i},D_{2i}) \right] & - \E\left[ Z_{2i} \right]
\end{bmatrix}
\equiv 
 \begin{bmatrix}
G_1 & - \E\left[Z_{1i} \right]\\
G_2 & - \E\left[Z_{2i} \right]
\end{bmatrix}
\end{align}
has rank $(d_\beta+1)$. 
\end{assumption}
\begin{assumption}\label{a:CLT}
A pointwise CLT applies: 
\begin{equation}\label{eqn:CLT}
\sqrt{n} \bigl\{ \hat{\vecf{M}}_n(\vecf{\beta}_{0},\tau_{0}) - \E[\hat{\vecf{M}}_n(\vecf{\beta}_{0},\tau_{0})] \bigr\}
\dconv  \Normalp{\vecf{0}}{\matf{\Sigma}} . 
\end{equation}
\end{assumption}
\begin{assumption}\label{a:G-est}
For $j=1,2$,
let $Z_{ji}^{(k)}$ denote the $k$th element of $\vecf{Z}_{ji}$, and similarly $\beta^{(k)}$. 
For $j=1,2$,
let $G_{j,kl}$ denote the row $k$, column $l$ element of $\matf{G_j}$ (from \cref{a:U}). 
Assume 
\begin{equation}\label{eqn:G-est}
-\frac{1}{nh_n} \sum_{i=1}^{n} 
 \tilde{I}'\bigl(-\Lambda(\vecf{Y}_{ji},\vecf{X}_{ji},\tilde{\vecf{\beta}})/h_n \bigr)
 Z_{ji}^{(k)} 
 \pD{}{\beta^{(l)}}\Lambda(\vecf{Y}_{ji},\vecf{X}_{ji},\vecf{\beta}) 
 \Bigr\rvert_{\vecf{\beta}=\tilde{\vecf{\beta}}} 
\pconv  G_{j,kl} 
 , 
\end{equation}
for each $k=1,\ldots,d_Z$ and $l=1,\ldots,d_\beta$, where $\tilde{\vecf{\beta}}$
lies between $\vecf{\beta}_{0}$ and $\hat{\vecf{\beta}}_{\mathrm{1s}}$ (defined in \cref{a:B} and \cref{eqn:def-est-1s}, respectively). 
\end{assumption}

\begin{assumption}\label{a:W}
For the weighting matrix, $\hat{\matf{W}} \pconv \matf{W}$, and both are symmetric, positive definite matrices. 
\end{assumption}

Our assumptions are an extended version of those in \citeposs{deCastroGalvaoKaplanLiu19}, as we use a similar smoothed GMM framework. 
The difference is that \citet{deCastroGalvaoKaplanLiu19} consider only a single choice at a fixed quantile. 
That is, they use instruments $Z_1$ (say for choice 1) to identify and estimate parameter $\beta$ at each given $\tau$.
In contrast, we consider multiple choices and jointly estimate the $\tau$ as well as the other parameters.
That is, we use instruments $Z_1$ for choice 1 and $Z_2$ for choice 2 ($d_{Z_1}+d_{Z_2}$ number of instruments in total) to jointly identify and estimate $(\beta,\tau)$.

The following is a brief discussion on our assumptions. 
\Cref{a:XYZ,a:Lambda,a:B,a:Z} extends \citeposs{deCastroGalvaoKaplanLiu19} assumptions in the single choice framework to our stack multiple choices framework. 
\Cref{a:B} assumes the population parameters are uniquely identified.
\Cref{a:Z} assumes the instruments have finite second moment. 
\Cref{a:Itilde,a:h} makes the same assumptions as in \citet{deCastroGalvaoKaplanLiu19} on the smoothing function (to replace the indicator function) and the smoothing bandwidth to make the smoothing asymptotically negligible (i.e., smoothing bandwidth goes to zero).
\Cref{a:Y} assumes the conditional distribution of the structural function given the instrument is continuous at zero.
\Cref{a:ULLN} assumes the ULLN of the smoothed moments.
%
%
\Cref{a:U} guarantees the local identification.
\Cref{a:CLT} assumes the CLT of the smoothed moments at the true parameter.
\Cref{a:G-est} guarantees the consistency of the asymptotic covariance matrix estimator.
\Cref{a:W} is a regular assumption for GMM. 
The detailed discussions and illustrations of the primitive conditions of the high-level assumption assumed in \cref{a:ULLN,a:CLT,a:G-est} can be found in \citet[p. 139--142]{deCastroGalvaoKaplanLiu19}.
For example, \citet[Lemma B.1]{deCastroGalvaoKaplanLiu19} provide sufficient lower-level assumptions that yields the ULLN assumed in \cref{a:ULLN}. Theorem 3.13 in \citet[Ch. 2]{Wooldridge86} and Proposition 1 of \cite{Andrews91CLT} provide conditions that yield the CLT in \cref{a:CLT} under weak dependence.

\subsection{Consistency}
\label{sec:asy-consistency}

In this section, we first show in \cref{lem:smooth-EMn-ULLN} that the sequence of (non-random) population of smoothed moments $\E[ \hat{\vecf{M}}_n(\vecf{\beta},\tau) ]$ converges to the unsmoothed population moment $\vecf{M}(\vecf{\beta},\tau)$, 
building on which we combine with \citet[Theorem 5.9]{vanderVaart98} to establish the consistency of the smoothed GMM estimator proposed in \cref{sec:est-GMM}. 
The proofs closely follow \citet{deCastroGalvaoKaplanLiu19}, but with extensions adapted to our framework, which stacks moments from multiple (at least two) choices and jointly estimates $\tau$ and other parameters $\beta$.

\begin{lemma}\label{lem:smooth-EMn-ULLN}
Under \Cref{a:XYZ,a:Lambda,a:B,a:Z,a:Y,a:Itilde,a:h}, 
using definitions in \cref{eqn:def-M,eqn:def-M-hat}, 
\begin{equation}\label{eqn:smooth-EMn-ULLN}
\sup_{(\beta, \tau)\in \mathcal{B}\times \mathcal{T} }
\absbig{ \E[ \hat{\vecf{M}}_n(\vecf{\beta},\tau) ] 
      - \vecf{M}(\vecf{\beta},\tau) }
= o(1) . 
\end{equation}
\end{lemma}

Let
\begin{equation}
\theta=(\beta\tr,\tau)\tr.
\end{equation}
\begin{theorem}\label{thm:consistency}
Under \Cref{a:XYZ,a:Lambda,a:B,a:Z,a:Y,a:Itilde,a:h,a:ULLN,a:W}, the smoothed GMM estimators from \cref{eqn:def-est-GMM} are consistent: 
\begin{equation}\label{eqn:consistency}
\hat{\theta}_{\mathrm{GMM}} - \theta_{0} = o_p(1) . 
\end{equation}
\end{theorem}

\subsection{Asymptotic normality}
\label{sec:asy-normality}

We first show a lemma that relies on \cref{a:CLT} and a proof that the asymptotic bias of smoothing is negligible, i.e., $\sqrt{n}\E[\hat{\vecf{M}}_n(\vecf{\beta}_{0},\tau_{0})] \to 0$. 
Building on this lemma, we establish the asymptotic normality of the smoothed GMM estimator.

\begin{lemma}\label{lem:Mn0-normality}
Under \Cref{a:XYZ,a:CLT,a:Itilde,a:Z,a:B,a:U,a:h,a:Lambda}, 
\begin{equation}\label{eqn:Mn0-normality}
\sqrt{n}\hat{M}_n(\beta_{0},\tau_{0})
\dconv 
\Normalp{\vecf{0}}{\matf{\Sigma}} , \quad 
\matf{\Sigma} 
= \lim_{n\to\infty} \Varp{ n^{-1/2}\sum_{i=1}^{n} \vecf{g}_{ni}(\vecf{\beta}_0,\tau_0) } . 
\end{equation}
With iid data and the conditional quantile restriction $\Pr\bigl( \Lambda(\vecf{Y}_{ji},\vecf{X}_{ji},\vecf{\beta}_{0})\le0 \mid \vecf{Z}_{ji}\bigr)=\tau_0$ for $j=1,2$, then 
$\matf{\Sigma} = \tau_0(1-\tau_0) \E(\vecf{Z}_i\vecf{Z}_i\tr)$.
\end{lemma}

\begin{theorem}\label{thm:normality}
Under \Cref{a:XYZ,a:Lambda,a:B,a:Z,a:Y,a:Itilde,a:h,a:ULLN,a:U,a:G-est,a:CLT,a:W}, the smoothed GMM estimator from \cref{eqn:def-est-GMM} has an asymptotic normal distribution
\begin{equation*}
\sqrt{n} ( \hat{\theta}_{\mathrm{GMM}} - \theta_{0} )
\dconv 
\Normal\bigl( \vecf{0} , (\matf{G}\tr \matf{W} \matf{G} )^{-1} 
\matf{G}\tr \matf{W} \matf{\Sigma} \matf{W} \matf{G}
(\matf{G}\tr \matf{W} \matf{G} )^{-1} \bigr) , 
\end{equation*}
where 
$\matf{G}$ is from \cref{eqn:def-G}, 
$\matf{W}$ is from \cref{a:W}, 
and 
$\matf{\Sigma}$ is from \cref{eqn:Mn0-normality}. 
\end{theorem}

When the weighting matrix is asymptotic efficient $\hat{\matf{W}} \pconv \matf{W} = \matf{\Sigma}^{-1}$,
the asymptotic covariance matrix in \cref{thm:normality} reduces to $(\matf{G}^{\tr} \matf{\Sigma}^{-1} \matf{G})^{-1}$
and the two-step GMM estimator defined in \cref{eqn:def-est-2s} is efficient. 
With iid data, the consistent estimator for $G$ and $\Sigma$ can be obtained as the sample average, as illustrated in \cref{sec:est-1s}.
For dependent data, there is a lack of a formal consistent long-run variance estimator in the literature.
A potential route is to apply \cite{deJongDavidson00}, which allows triangular arrays and weak dependence. 
We leave this question for future work.

The results in this section establish the consistency and asymptotic normality of the proposed smooth GMM estimator with standard root-$n$ convergence rate. This latter result is in contrast with the cubic-root rate in \citet{deCastroGalvaoOta24}. 
Our $\sqrt{n}$ rate comes from using the GMM method with smoothed moment functions (see \citet{Hansen82}, \citet{NeweyMcFadden94}, \citet{HallInoue03}, and \citet{KangLeeSong25}).
 

\section{Application}\label{sec:Application}

In this section, we first employ our general quantile Euler equation \eqref{eq:Euler suming dimensions} to an intertemporal consumption model with multiple assets. 
Then, we apply our method to estimate the main parameters characterizing the preferences, i.e., the risk attitude and the elasticity of intertemporal substitution.

\subsection{Intertemporal consumption model with multiple assets} \label{sec:assets_agents}

Consider that consumption and investment take place at a discrete time point $t$. At the beginning of period $t$, the economic agent has wealth  $w_{t}$ for consumption and investment.

There are $m$ assets (Lucas' trees), that give fruit each period.
The price and dividend of asset $j\in \{1,...,m\}$ are determined by measurable functions $p_{j,t}: \mathcal{E} \to \R_{+}$ and $d_{j,t}: \mathcal{E} \to \R_{+}$, respectively. 
Given  price $p_{j,t}=p_{j,t}(e_{t})$ and dividend $d_{j,t}=d_{j,t}(e_{t})\in \R_{+}$, asset $j$'s return from  $t-1$ to   $t$ is   $R_{j,t}=\frac{p_{j,t}+d_{j,t}}{p_{j,t-1}}$, where we omit the dependence on the shock $e_{t}$.  
The following vectorial notation will be useful below:
\begin{align*}
    p_{t} =\left(p_{1,t},\ldots, p_{m,t}  \right) \; ; \;
     d_{t} =\left(d_{1,t},\ldots, d_{m,t}  \right) \; ; \;
    \text{ and }  
     R_{t}=\left( 
R_{1,t},\ldots,
R_{j,t},\ldots,
R_{m,t} \right).
\end{align*}
 Let $a_{t}=\left( a_{1,t},...,a_{m,t} \right)$ be the $m$-vector of assets purchased at  the end of period $t-1$ and held until period $t$.
Thus, $x_{t}=  p_{t-1} 
    \cdot a_{t} $ denotes the investment by the economic agent in period $t-1$. 
At the beginning of period $t$, the individual has wealth $w_{t}=(p_{t}+d_{t})\cdot a_{t}$. From this,  the return $R_{t}$ of agent's  portfolio is 
\begin{align*}
    R_{t}=\frac{w_{t}}{x_{t}}= \frac{ \left( p_{t}+ d_{t} \right)  
    \cdot a_{t} }{p_{t-1} \cdot a_{t} }, 
\end{align*}
so that $w_{t}=x_{t}R_{t}$. 
Let $\omega_{j,t}$ denote the weight of asset $j$  in the value of the portfolio  bought in  $t-1$ and  held until $t$, that is, 
\begin{align}\label{eq:omega_j_t}
\omega_{j,t}=\frac{p_{j,t-1}\, a_{j,t}}{p_{t-1}\cdot a_{t}} . 
\end{align}
It is easy to see that these weights sum up to 1, that is, $\sum_{j=1}^{m}\omega_{j,t}=1$. 
We can write $R_{t}$ as 
\begin{align}
\nonumber
R_{t} &=  \frac{(p_{t}+d_{t})\cdot a_{t}}
{p_{t-1}\cdot a_{t}}= \frac{\sum_{j=1}^{m} (p_{j,t}+d_{j,t}) a_{j,t}}
{p_{t-1}\cdot a_{t}} = \sum_{j=1}^m \left( \frac{p_{j,t}+d_{j,t}}{p_{j,t-1}} \right)  \cdot 
\left( \frac{p_{j,t-1}\, a_{j,t}}{p_{t-1}\cdot a_{t}} \right)\\ 
&=\sum_{j=1}^m R_{j,t} \cdot \omega_{j,t}.
\label{eq:zti-from-omegajti}
\end{align}

Notice that given the value saved at the end of period $t-1$,  $x_{t}=p_{t-1}\cdot a_{t}$,
and the weight $\omega_{j,t}$ of asset $j$ in period $t$, we can obtain 
\begin{align*}
    a_{j,t}& = x_{t}   \omega_{j,t}
    \Rightarrow
    a_{t}=\left( a_{1,t},...,a_{m,t} \right)= x_{t}  \omega_{t} =  x_{t} \left( \omega_{1,t},  \ldots, \omega_{m,t} \right) .
\end{align*}
Therefore, we can separate the agent's problem in two parts: agent decides how much to invest (or save), $x_{t}$, and, separately, the portfolio weights $\omega_{j,t}$ of each asset. 
Once the weights $\omega_{j,t}$ are defined, we obtain $R_{t}$
by \eqref{eq:zti-from-omegajti} and we can define the savings problem taking this $R_{t}$ as given. 
Below, we focus first on the savings problem, taking the portfolio as given. 
Later, we return to the portfolio problem.
Given  wealth $w_{t}=x_{t}R_{t}$, agent chooses to save $x_{t+1}$ and consume $c_{t}=w_{t} - x_{t+1}$. 
Once the decision-maker (DM) chooses a portfolio $\omega_t$, in time $t$, its return will be simply $R(e_{t+1})\cdot\omega_t$.  
The budget constraint can be written as follows
\begin{equation}
\label{eq:xt+1}
    x_{t+1} = (x_t - c_t)R(e_{t+1})\cdot \omega_t ,
\end{equation}
where wealth tomorrow is equal to the net wealth today (from subtracting current consumption) times the return of the portfolio.

The general dynamic model in \cref{eq:FE} can be specialized to a problem to choose a sequence with both consumption and portfolio weights, $\left(c_{t},\omega_t\right)_{t\geq0}$, to maximize the following recursive equation: 
\begin{align}\label{eq:def_V}
V( x_{t}, e_{t}) &=    
\max_{\substack{c_{t} \in [0,x_t], \\
\omega_t \in \Delta_m}}
\ \bigl\{  U(c_{t}) + \delta \Q \left[ V(x_{t+1},e_{t+1}) | \,  \mathcal{F}_{t} \right] \bigr\} \\
& \textnormal{s.t.} \quad x_{t+1} = (x_t - c_t)R(e_{t+1})\cdot \omega_t ,
\end{align}
where $U(c_{t})$ is the instantaneous utility function, $\delta \in (0,1)$ is the discount factor, and $\Q[\cdot]$ is the conditional quantile function for a given risk attitude $\tau$. In this model, the quantile $\tau$ captures the risk attitude (see Appendix \ref{app:preferences} for further discussion on risk attitude). 
In the quantile model in equation \eqref{eq:def_V} the DM considering how much to invest in risky assets chooses the policy ensuring the highest $\tau$-quantile.
This justifies the assertion that quantile-maximizers are concerned with downside risk (possibility of losses below some critical level). 

We focus on a common isoelastic type of utility function, which is defined by:
\begin{align}\label{eq:isoelastic_function}
    U(c) = 
   \frac{1}{1-\gamma} c^{1-\gamma},  \;\quad   \text{ with } 
   \gamma>0,\gamma\not = 1 . 
\end{align}
Moreover, assume that the return vector $R_t$ depends on the shock $e_t$ but is time invariant, that is, $R_t \equiv R$.

Then, the dynamic problem above takes the following form
\begin{align}
\label{eq:multi asset problem}
V(x_{t}, e_{t}) &=    
\max_{\substack{c_{t} \in [0,x_t], \\
\omega_t \in \Delta_m}}
\ \left\{ \frac{c_{t}^{1-\gamma}}{1-\gamma} + \delta \Q \left[ V((x_t-c_t)R(e_{t+1})\cdot \omega_t,e_{t+1}) | \,  \mathcal{F}_{t} \right] \right\} .
\end{align}
The recursive model in \eqref{eq:multi asset problem} has three parameters characterizing the preferences: $\delta$, $\tau$, and $\gamma$. The first parameter, $\delta$, describes the intertemporal discount. As mentioned previously, the $\tau$ quantile captures risk attitude. The latter captures the EIS as $1/\gamma$. 
One advantage of the quantile model is that it allows for a complete separation between the risk attitude, $\tau$, and the EIS, $1/\gamma$ (see, Appendix \ref{app:preferences} for a discussion on this separation).

The value function in \eqref{eq:multi asset problem} has a fixed point. 
In addition, the quantile Euler equation for this dynamic consumption model is summarized in the following result.

\begin{lemma}[de Castro, Galvao,  and Ota (2026)]\label{theorem:QP}
Let Assumptions \ref{ass:basic} hold. 
Assume that $R: \mathcal{E} \to \R_{+}^M$ is continuous and bounded and $U: \R_{+} \to \R$ is given by \eqref{eq:isoelastic_function}.  
Let $\bar{\omega}$ denote the optimal portfolio that solves \eqref{eq:multi asset problem}.  
Then, the following Euler Equation holds:
\begin{align}\label{eq:asset combined gamma01}
 &  \Q\left[\delta\left(\frac{c_{t+1}}{c_t}\right)^{-\gamma} 
  R_j (e_{t+1}) \Big| \,  \mathcal{F}_{t} \right] = 1,
\end{align}
for any asset $j$ in the portfolio such that $\bar{\omega}_{tj}>0$.
\end{lemma}

The Euler equation \eqref{eq:asset combined gamma01} has a very simple interpretation where the marginal utility of consumption between the two periods must be equal. 
Using the previous notation, 
for the case of two returns $j=1,2$, the structural function is $\Lambda=\delta \left(\frac{c_{t+1}}{c_t}\right)^{-\gamma} R_{j}$ and the parameters of interest are $(\delta,\gamma,\tau)$.

\subsection{Estimating equation}

We are interested in estimating the parameters characterizing the Euler equation in \cref{eq:asset combined gamma01}. For a given asset, it can be written as
\begin{equation}
   \Q\left[\delta\left(\frac{c_{t+1}}{c_t}\right)^{-\gamma} R (e_{t+1})  \Big| \,  \mathcal{F}_{t} \right] = 1,
\end{equation}
which are: $\tau$, $\delta$, and $\gamma$.
The quantile operator has an equivariance (with respect to monotone transformations) property, such that $\Q[\ln(W)]=\ln(\Q[W])$.
Then, we can log-linearize, without errors, the quantile Euler equation \cref{eq:asset combined gamma01} as
\begin{equation}
   \Q\left[\ln\delta-\gamma\ln (c_{t+1}/c_{t}) +\ln R (e_{t+1})  \Big| \,  \mathcal{F}_{t} \right] = 0.
\end{equation}
Using the fact that $\Q[W]=-Q_{1-\tau}[-W]$, when $\gamma>0$, the previous equation reduces to 
\begin{equation}
     \textnormal{Q}_{1-\tau}\left[\ln(c_{t+1}/c_{t})-\gamma^{-1}\ln\delta-\gamma^{-1}\ln R (e_{t+1})  \Big| \,  \mathcal{F}_{t} \right] = 0.
\end{equation}
where $\gamma^{-1}\ln\delta$ and $\gamma^{-1}$ are the intercept and slope parameter when performing a $(1-\tau)$ IV quantile regression of $\ln(c_{t+1}/c_{t})$ on $\ln R (e_{t+1})$ with instruments $z_t$.


\subsection{Data}

We use the dataset from \cite{arellano2017earnings}, which is originally from the Panel Study of Income Dynamics (PSID). 
It contains the household consumption expenditure of nondurable goods and services and asset holdings from 1999 to 2009 (six time periods). 
\cite{arellano2017earnings} study the earnings and consumption model. 
We use the data to study our quantile Euler equation.
Following \citet{Yogo04,deCastroGalvao19},
we use the consumption growth rate ($\ln(c_{t+1}/c_{t})$) as the outcome variable. 
We use the price level ($1+\textnormal{inf}_t=P_{t+1}/P_t$) to compute the inflation rate.
We use the assets adjusted for price level as the real assets and the log of the real asset growth rate as the real return rate, which is our endogenous variable.\footnote{Real asset: $a_{t}=A_{t}/P_{t}$. Real return rate: $\ln(1+r_{t})=\ln(a_{t+1}/a_{t})$}
We consider two assets. One is the household financial assets, which include cash, bonds, stocks, business values, and pension funds, net of other debt.\footnote{Variable name: \code{fin_assets3} in \citet{arellano2017earnings}.} 
The other asset is the household's total assets, which are the sum of financial assets, real-estate values, and car values, net of mortgages and other debt.\footnote{Variable name: \code{tot_assets3} in \citet{arellano2017earnings}.}
Moreover, we use the twice-lagged consumption growth rate, the twice-lagged nominal asset return rate,\footnote{For the nominal asset return, we use the log of asset growth rate $\ln(A_{t+1}/A_{t})$.} and the inflation rate as instruments, 
similar to the instruments used in \cite{Yogo04,deCastroGalvao19,deCastroGalvaoKaplanLiu19}.

Similar to \cite{arellano2017earnings}, we focus on households with heads aged between 25 and 60; we drop the observations whose consumption or asset either in levels or in logs are missing; and we focus on a balanced sample of $N=525$ households in three time periods from 2005 to 2009.

\subsection{Results}

The estimation results are described in \cref{tab:emp:bs} with smoothing bandwidth from \citet{KaplanSun17}. 
It reports the estimates and corresponding confidence intervals for the risk attitude / quantile level ($\tau$), the discounting factor ($\delta$), and $\gamma$ as well as the EIS ($1/\gamma$). 
The estimated $\tau$ is $0.402$.
The fact that the estimated risk attitude ($\tau$) is less than the median indicates some degree of risk aversion.
The estimated discounting factor is 1.197, which is around 1.
The estimated EIS is about 0.83, which is consistent with the literature (see, e.g., \citet{Havranek15}, \citet{Thimme17}, and \citet{Bestetal20}).


\begin{table}[htbp] 
    \centering
\caption{\label{tab:emp:bs} Estimates of parameters.}
\begin{threeparttable}
\begin{tabular}[c]{S[table-format=4.0,round-precision=0] 
    cS[table-format=1.3,round-precision=3]S[table-format=1.3,round-precision=3]
    S[table-format=1.3,round-precision=3]
    >{{[}} 
    S[table-format=4.3,round-precision=3]
    @{,\,} 
    S[table-format=2.3,round-precision=3]
    <{{]}} 
  }
  \toprule
       \multicolumn{1}{c}{Parameters}          && {Estimates} & {SE}  && \multicolumn{2}{c}{95\% CI} \\
\midrule
$\tau$           &&    0.402151  &    0.030903  &&    0.341581  &    0.462721     \\
$\delta$          &&    1.197456  &    0.075172  &&    1.050118  &    1.344794     \\
$\gamma$         &&    1.205733  &    0.473464  &&    0.277744  &    2.133722     \\
{EIS}            &&    0.829371  &    0.248475  &&    0.342360  &    1.316382     \\
\bottomrule
\end{tabular}
\begin{tablenotes}
\item Following \cite{ChernozhukovFernandezValMelly13}, we obtain the bootstrapped standard error through 1000 bootstraps. We use estimates $\pm$ 1.96 times the bootstrapped standard error to obtain a 95\% CI. We use \cite{KaplanSun17} plug-in smoothing bandwidth, which ranges from 0.4 to 1.6 with mean value around 0.7.
\end{tablenotes}
\end{threeparttable}
\end{table}

We use bootstrapped standard errors to construct confidence intervals.
The bootstrapped procedure used captures well the data structure, that is, it is iid across households and time-dependent within households. 
The bootstrap consistency is shown in \cite{ChernozhukovFernandezValMelly13}.
We leave the task of establishing statistical properties of the standard error in a panel data framework to future research.

\cref{tab:emp:bs:allh} in the Appendix reports estimates for robustness results using other smoothing bandwidths,
which suggests the effect of smoothing is very limited.
The EIS point estimates are very close using various bandwidths, ranging from 0.69 to 0.83.
Using other (small) bandwidths, the discounting factor point estimates are all around 1, ranging from 1.17 to 1.20. 
The estimated risk attitude is near the median from 0.402
to 0.414. 
The confidence intervals for the parameters are very similar across the bandwidths.
For example, the CI for EIS is around (0, 1.5) and the CI for risk attitude $\tau$ is around (0.3, 0.5) when using bandwidth $h=\{0.01, 0.1\}$.

\section{Monte Carlo}\label{sec:MC}

In this section, we illustrate the finite-sample performance of our proposed methods across different data-generating processes (DGP). We evaluate the empirical bias and the root mean squared error (RMSE) of the estimator.

\subsection{Design}

We consider two different DGPs that allow for unobserved heterogeneity and endogeneity.

DGP 1 considers generating two processes for the dependent variable, $Y$. 
We generate dataset 1 (or asset 1) from
\begin{align}
    Y_{1i} & =\beta_{01}(U_{1i})+\beta_{11}(U_{1i})D_{1i}, 
    U_{1i}  \stackrel{\mathit{iid}}\sim \UnifDist (0,1), 
    D_{1i}  = Z_{1i} +U_{1i},
    Z_{1i}  \stackrel{\mathit{iid}}\sim N(4,1), \label{eqn:sim:asset1}\\
    \beta_{01}&(u) = \Phi^{-1}(u), \ \beta_{11}(u)= 0.7,
\end{align}
where $\Phi(\cdot)$ is the CDF of $N(0,1)$, and  $\Phi^{-1}(\cdot)$ is the quantile function of $N(0,1)$. Also, we generate dataset 2 (or asset 2) from the processes
\begin{align}
    Y_{2i} & =\beta_{02}(U_{2i})+\beta_{12}(U_{2i})D_{2i},
    U_{2i}  \stackrel{\mathit{iid}}\sim \UnifDist (0,1), 
    D_{2i}  = Z_{2i} +U_{2i}, 
    Z_{2i}  \stackrel{\mathit{iid}}\sim N(4,1), \label{eqn:sim:asset2}\\
    \beta_{02}&(u) =\Phi^{-1}(0.7), \beta_{12}(u)= u.
\end{align}

In both processes, we allow endogeneity in regressors. That is, the variable $D_{1i}$ and innovation $U_{1i}$ are correlated, and in the same way that $D_{2i}$ and $U_{2i}$ are correlated.
We use instrument variables ($Z_{1i}$ and $Z_{2i}$ respectively) to solve the endogeneity.

In this first design, the structural quantile functions (SQF) of the two assets overlap at quantile $\tau=0.7$. 
That is, the $(\beta_{01}(\tau), \beta_{11}(\tau))$ and $(\beta_{02}(\tau), \beta_{12}(\tau))$ match only at quantile $\tau=0.7$.

We consider a second model.
DGP 2 is identical to DGP 1 in equations \cref{eqn:sim:asset1,eqn:sim:asset2} above except that we modify the random coefficients as following
\begin{align}
         \beta_{01}&(u) =5u , \beta_{11}(u)= 5u,\\
     \beta_{02}&(u) =6u-0.7 , \beta_{12}(u)= 6u-0.7.
\end{align}
In DGP 2, the two assets' SQFs overlap exactly at quantile $\tau=0.7$, although the two assets' SQFs are very close at other quantiles.

\Cref{fig:sim:all}, in \cref{sec:MCgraph}, illustrates the two assets' SQFs at various quantiles and the coefficient as a function of the quantile level. 
Compared to DGP 1, where the two assets have very distinct SQFs, the two assets in DGP 2 have very close SQFs, and hence it is more challenging in DGP 2 to estimate the quantile levels at which both assets' SQFs meet exactly.

\subsection{Results}

Now we present results for the empirical bias and RMSE for the three parameters of interest, $(\tau,\beta_{1},\beta_{0})$. We consider three sample sizes $n\in\{1500,3000,5000\}$. The number of replications is 1000.

\begin{table}[htbp] 
    \centering\caption{\label{tab:sim:DGP1-2} Percentage bias and RMSE of parameters in simulation.}\begin{threeparttable}
\begin{tabular}[c]{cS[table-format=4.0,round-precision=0] 
  cS[table-format=2.1,round-precision=1]S[table-format=1.3,round-precision=3]
  cS[table-format=2.1,round-precision=1]S[table-format=1.3,round-precision=3]
  cS[table-format=2.1,round-precision=1]S[table-format=1.3,round-precision=3]
  }
  \toprule
  &   &&  \multicolumn{2}{c}{$\hat{\tau}$}          && \multicolumn{2}{c}{$\hat{\beta}_1$} && \multicolumn{2}{c}{$\hat{\beta}_0$} \\
  \cmidrule{4-5} \cmidrule{7-8} \cmidrule{10-11}
{DGP} & \multicolumn{1}{c}{$n$} &&  {\% Bias} & {RMSE} &&  {\% Bias} & {RMSE} && {\% Bias} & {RMSE}  \\
  \midrule
1 & 1500        &&    1.864429  &    0.043310  &&    0.815714  &    0.027974  &&    5.214716  &    0.153422     \\
  & 3000        &&    0.651714  &    0.026052  &&    0.229571  &    0.018712  &&    2.562927  &    0.105266     \\
  & 5000        &&    0.385143  &    0.019528  &&    0.185429  &    0.014722  &&    1.326276  &    0.080348     \\
\midrule
2 & 1500        &&   -1.022857  &    0.088957  &&   -0.746143  &    0.533015  &&   -4.100143  &    1.022516     \\
  & 3000        &&   -0.971857  &    0.063449  &&   -0.745343  &    0.380732  &&   -3.214171  &    0.766956     \\
  & 5000        &&   -0.798143  &    0.048720  &&   -0.497829  &    0.295990  &&   -2.990457  &    0.598567     \\

\bottomrule
\end{tabular}
\begin{tablenotes}
\item 1000 replications. \citet{KaplanSun17} plug-in bandwidth is used.
Percentage bias is computed as the bias divided by the true parameter, then multiplied by 100. 
\end{tablenotes}
\end{threeparttable}
\end{table}

\cref{tab:sim:DGP1-2} reports the percentage bias and RMSE of the smooth GMM estimator proposed in this paper using both the DGPs. 
The smoothing bandwidth is from \citeposs{KaplanSun17} plug-in bandwidth, which ranges from 
0.92 to 2.61 in DGP 1 and from 6.0 to 37.7 in DGP 2. 
We report the simulation results with other small bandwidths 
in \cref{tab:sim:DGP:all} in the Appendix.
The results patterns are similar, indicating the estimation is robust to the smoothing bandwidth, when the smoothing is small.

In \cref{tab:sim:DGP1-2}, all the parameters are estimated well with small percentage bias.
The percentage bias of $\tau$ is less than 2\%.
The corresponding absolute value of bias of $\tau$ is less than 0.01. 
The bias and RMSE decrease towards zero as we increase the sample size. 
The estimates in DGP 2 have a relatively larger percentage bias and RMSE than in DGP 1, which reflects the challenges of estimation in DGP 2.
Still, the estimates in DGP 2 have a very small percentage bias with larger sample sizes. 
For example, it has 0.8\% bias (0.006 in absolute value) in $\tau$ and up to 3\% bias in other coefficients with $n=5000$. Overall, the simulations provide numerical evidence that the proposed methods perform well in finite samples.

\section{Conclusion}\label{sec:Conclusion}

This paper develops estimation procedures for all the parameters in a dynamic quantile preferences model including the quantile, which captures the risk attitude of the economic agent. Identification of the quantile of interest relies on the availability of at least two assets. We propose a smooth generalized method of moments for estimation, and provide conditions for the estimator to be consistent and asymptotically normal. Monte Carlo simulations provide numerical evidence that the estimator possesses good finite sample property. An empirical application to intertemporal consumption documents evidence of risk aversion, and EIS relatively close to one.

Many issues remain to be investigated. As in most problems involving instrumental variables, the issue of potential weak instruments poses challenging new questions. Optimal selection of nuisance parameters is also a critical direction for future research.





\bibliographystyle{elsarticle-harv} 
\bibliography{nlqriv}


\newpage

\setcounter{page}{1}
\appendix

\begin{appendices}

This Online Supplemental Appendix provides several additional results. 
Section \ref{sec:app-proofs} collects the proofs of the results in the main text. Section \ref{sec:bandwidth} discusses the sensitivity to bandwidth. Additional details for Monte Carlo are provided in Section \ref{sec:MCgraph}. Finally, Section \ref{app:preferences} of this appendix reviews properties of quantiles, defines quantile preferences, and discusses the separation of risk and elasticity of intertemporal substitution.

\section{Proofs}
\label{sec:app-proofs}



\begin{proof}[Proof of \Cref{lem:smooth-EMn-ULLN}]
This proof follows closely to the proof of Lemma 2 in \citet{deCastroGalvaoKaplanLiu19}.
Noting that $\absbig{\vecf{Z}[\tilde{I}(\cdot)-\Ind{\cdot}]} \le \abs{\vecf{Z}}$ (i.e., $\abs{\vecf{Z}}$ is a dominating function) and applying the dominated convergence theorem (since $\vecf{Z}$ has finite expectation by \cref{a:Z}), since $h_n\to0$ by \cref{a:h}, 
\begin{align*}
& \lim_{h_n\to0} \sup_{(\beta, \tau)\in \mathcal{B}\times \mathcal{T}
} 
\normbig{  \E[ \hat{\vecf{M}}_n(\vecf{\beta},\tau) ] 
       -\E\left[\vecf{Z}\left(\Ind{\Lambda(\vecf{Y},\vecf{X},\vecf{\beta})\le0}-\tau\right) \right]
}
\\&= \lim_{h_n\to0} \max_{\vecf{\beta}\in\mathcal{B}} 
\normbig{ \E\left\{\vecf{Z}\left[\tilde{I}\left(\frac{-\Lambda(\vecf{Y},\vecf{X},\vecf{\beta})}{h_n}\right) - \Ind{\Lambda(\vecf{Y},\vecf{X},\vecf{\beta})\le0}\right] \right\}
}
\\&= \lim_{h_n\to0} 
\normbig{ \E\left\{\vecf{Z}\left[\tilde{I}\left(\frac{-\Lambda(\vecf{Y},\vecf{X},\vecf{\beta}^*_n)}{h_n}\right) - \Ind{\Lambda(\vecf{Y},\vecf{X},\vecf{\beta}^*_n)\le0}\right] \right\}
}
\\&= \normbig{ \E\left\{
      \lim_{h_n\to0} \vecf{Z} \left[\tilde{I}\left(\frac{-\Lambda(\vecf{Y},\vecf{X},\vecf{\beta}^*_n)}{h_n}\right) - \Ind{\Lambda(\vecf{Y},\vecf{X},\vecf{\beta}^*_n)\le0}\right] 
   \right\}
}
\\&= \vecf{0} 
\label{eqn:smooth-unsmooth-limit-DCT}\refstepcounter{equation}\tag{\theequation}
\end{align*}
as long as there is no probability mass at $\Lambda(\vecf{Y},\vecf{X},\vecf{\beta})=0$ for any $\vecf{\beta}\in\mathcal{B}$ and almost all $\vecf{Z}$, which is indeed true by \Cref{a:Y}. 
The notation $\vecf{\beta}^*_n$ denotes the value attaining the maximum, which exists since $\mathcal{B}$ is compact by \cref{a:B}. 
\end{proof}

\begin{proof}[Proof of \Cref{thm:consistency}]
This proof follows closely to the Theorem 3 of \citet{deCastroGalvaoKaplanLiu19}.
We extend \citeposs{deCastroGalvaoKaplanLiu19} proof to adapt for our framework of stacked moments and joint estimation of $\tau$ and other parameters.

To prove the consistency of $\hat{\vecf{\theta}}_{\mathrm{GMM}}=(\hat{\vecf{\beta}}_{\mathrm{GMM}},\hat{\tau}_{\mathrm{GMM}})$, we show that the two conditions of Theorem 5.7 in \citet{vanderVaart98} are satisfied. 
The proof follows closely to the proof of \citeposs{deCastroGalvaoKaplanLiu19} Theorem 3, but is extended to fit in with our settings and assumptions.
The first condition of Theorem 5.7 in \citet{vanderVaart98} requires 
\begin{equation}\label{eqn:Thm5.7-cond1}
\sup_{(\beta, \tau)\in \mathcal{B}\times \mathcal{T} } 
\abs{ \hat{\vecf{M}}_n\left(\vecf{\beta},\tau\right)\tr 
\hat{\matf{W} }   \hat{\vecf{M}}_n\left(\vecf{\beta},\tau\right) -\vecf{M}\left(\vecf{\beta},\tau\right)\tr  
\matf{W}
\vecf{M}\left(\vecf{\beta},\tau\right) }
\pconv 0.
\end{equation}

Combining results from \cref{a:ULLN,lem:smooth-EMn-ULLN} and the triangle inequality,
\begin{align*}
& \sup_{
(\beta, \tau)\in \mathcal{B}\times \mathcal{T}
}
\absbig{ \hat{\vecf{M}}_n(\vecf{\beta},\tau) - 
\vecf{M}(\vecf{\beta},\tau)
}
\\&=
\sup_{
(\beta, \tau)\in \mathcal{B}\times \mathcal{T}
}
\absbig{ \hat{\vecf{M}}_n(\vecf{\beta},\tau) 
    - \E\bigl[ \hat{\vecf{M}}_n(\vecf{\beta},\tau) \bigr] 
    + \E\bigl[ \hat{\vecf{M}}_n(\vecf{\beta},\tau) \bigr] 
    - \vecf{M}(\vecf{\beta},\tau) }
\\&\le
\overbrace{\sup_{
(\beta, \tau)\in \mathcal{B}\times \mathcal{T}
}
\absbig{ \hat{\vecf{M}}_n(\vecf{\beta},\tau) 
    - \E\bigl[ \hat{\vecf{M}}_n(\vecf{\beta},\tau) \bigr] }}^{=o_p(1)\textrm{ by \cref{a:ULLN}}}
+
\overbrace{\sup_{
(\beta, \tau)\in \mathcal{B}\times \mathcal{T}
}
\absbig{ \E\bigl[ \hat{\vecf{M}}_n(\vecf{\beta},\tau) \bigr] 
    - \vecf{M}(\vecf{\beta},\tau) }}^{=o_p(1)\textrm{ by \cref{lem:smooth-EMn-ULLN}}}
\\&= o_p(1)+o_p(1)
   = o_p(1) . 
\label{eqn:Mn-pt-Thm59-cond1}\refstepcounter{equation}\tag{\theequation}
\end{align*}

From \cref{eqn:Mn-pt-Thm59-cond1}, 
$\sup_{(\beta, \tau)\in \mathcal{B}\times \mathcal{T}}
 \norm{ \hat{\vecf{M}}_n(\vecf{\beta},\tau) - 
\vecf{M}(\vecf{\beta},\tau)
}  = o_p(1)$. 
From \cref{a:W}, $ \hat{\matf{W}} = \matf{W} + o_p(1) $, which does not depend on $\vecf{\beta}$. 

Let $\norm{\cdot}$ denote the Frobenius matrix norm 
$\norm{\matf{A}}=\norm{\matf{A}\tr}=\sqrt{\textrm{tr}(\matf{A}\matf{A}\tr)}$, 
which is the Euclidean norm if $\matf{A}$ is a vector. 
Given this norm, the Cauchy--Schwarz inequality states that for any matrices $\matf{A}$ and $\matf{B}$,  $\norm{\matf{A}\matf{B} } \le \norm{\matf{A}} \norm{\matf{B}} $. 

We now use the triangle inequality, Cauchy--Schwarz inequality, uniform convergence in probability of $\hat{\vecf{M}}_n(\vecf{\beta},\tau)$, and convergence in probability of $\hat{\matf{W}}$, to show the required condition in \cref{eqn:Thm5.7-cond1}: 
\begin{align*}
& \sup_{(\beta, \tau)\in \mathcal{B}\times \mathcal{T}} 
\bigl\lvert
 \hat{\vecf{M}}_n(\vecf{\beta},\tau)\tr  \hat{\matf{W} }   
      \hat{\vecf{M}}_n(\vecf{\beta},\tau) 
     -\vecf{M}(\vecf{\beta},\tau)\tr  \matf{W}   
      \vecf{M}(\vecf{\beta},\tau) 
\bigr\rvert 
\\&= \sup_{(\beta, \tau)\in \mathcal{B}\times \mathcal{T}} 
\bigl\lvert
(  \vecf{M}(\vecf{\beta},\tau)
 +(\hat{\vecf{M}}(\vecf{\beta},\tau)-\vecf{M}(\vecf{\beta},\tau)  ))\tr 
(\matf{W} + (\hat{\matf{W}}-\matf{W})) 
\\&\qquad\quad\times
(  \vecf{M}(\vecf{\beta},\tau)
 +(\hat{\vecf{M}}(\vecf{\beta},\tau) - \vecf{M}(\vecf{\beta},\tau)  )) 
\\&\qquad\quad
-\vecf{M}(\vecf{\beta},\tau)\tr  \matf{W}  \vecf{M}(\vecf{\beta},\tau) 
\bigr\rvert 
\\&= \sup_{(\beta, \tau)\in \mathcal{B}\times \mathcal{T}}  
\bigl\lvert
   \vecf{M}(\vecf{\beta},\tau)\tr 
   (\hat{\matf{W}}-\matf{W})    
   \vecf{M}(\vecf{\beta},\tau)
+
  \vecf{M}(\vecf{\beta},\tau)\tr 
  \matf{W} 
  (\hat{\vecf{M}}(\vecf{\beta},\tau)-\vecf{M}(\vecf{\beta},\tau)  )
\\&\quad 
+ 
  (\hat{\vecf{M}}(\vecf{\beta},\tau)-\vecf{M}(\vecf{\beta},\tau) )\tr
  \matf{W}
  \vecf{M}(\vecf{\beta},\tau) 
+ 
  \vecf{M}(\vecf{\beta},\tau)\tr 
  (\hat{\matf{W}}-\matf{W}) 
  (\hat{\vecf{M}}(\vecf{\beta},\tau)-\vecf{M}(\vecf{\beta},\tau) )
\\&\quad + 
  (\hat{\vecf{M}}(\vecf{\beta},\tau)-\vecf{M}(\vecf{\beta},\tau) )\tr
  \matf{W} 
  (\hat{\vecf{M}}(\vecf{\beta},\tau)-\vecf{M}(\vecf{\beta},\tau) )
\\&\quad + 
   (\hat{\vecf{M}}(\vecf{\beta},\tau)-\vecf{M}(\vecf{\beta},\tau) )\tr
   (\hat{\matf{W}}-\matf{W})    
   \vecf{M}(\vecf{\beta},\tau)
\\
&\quad + 
   (\hat{\vecf{M}}(\vecf{\beta},\tau)-\vecf{M}(\vecf{\beta},\tau) )\tr
   (\hat{\matf{W}}-\matf{W})    
   (\hat{\vecf{M}}(\vecf{\beta},\tau)-\vecf{M}(\vecf{\beta},\tau) )
\bigr\rvert \\
&\le \sup_{(\beta, \tau)\in \mathcal{B}\times \mathcal{T}}  
\abs{ 
   \vecf{M}(\vecf{\beta},\tau)\tr 
   (\hat{\matf{W}}-\matf{W})    
   \vecf{M}(\vecf{\beta},\tau)
} 
+ \sup_{(\beta, \tau)\in \mathcal{B}\times \mathcal{T}} 
\abs{
  \vecf{M}(\vecf{\beta},\tau)\tr 
  \matf{W} 
  (\hat{\vecf{M}}(\vecf{\beta},\tau)-\vecf{M}(\vecf{\beta},\tau)  )
} \\
&\quad + \sup_{(\beta, \tau)\in \mathcal{B}\times \mathcal{T}} 
\abs{
  (\hat{\vecf{M}}(\vecf{\beta},\tau)-\vecf{M}(\vecf{\beta},\tau) )\tr
  \matf{W}
  \vecf{M}(\vecf{\beta},\tau) 
} 
+ \sup_{(\beta, \tau)\in \mathcal{B}\times \mathcal{T}} 
\abs{ 
  \vecf{M}(\vecf{\beta},\tau)\tr 
  (\hat{\matf{W}}-\matf{W}) 
  (\hat{\vecf{M}}(\vecf{\beta},\tau)-\vecf{M}(\vecf{\beta},\tau) )
}\\
&\quad + \sup_{(\beta, \tau)\in \mathcal{B}\times \mathcal{T}} 
\abs{
  (\hat{\vecf{M}}(\vecf{\beta},\tau)-\vecf{M}(\vecf{\beta},\tau) )\tr
  \matf{W} 
  (\hat{\vecf{M}}(\vecf{\beta},\tau)-\vecf{M}(\vecf{\beta},\tau) )
}\\
&\quad + \sup_{(\beta, \tau)\in \mathcal{B}\times \mathcal{T}} 
\abs{
   (\hat{\vecf{M}}(\vecf{\beta},\tau)-\vecf{M}(\vecf{\beta},\tau) )\tr
   (\hat{\matf{W}}-\matf{W})    
   \vecf{M}(\vecf{\beta},\tau)
}\\
&\quad + \sup_{(\beta, \tau)\in \mathcal{B}\times \mathcal{T}} 
\abs{
   (\hat{\vecf{M}}(\vecf{\beta},\tau)-\vecf{M}(\vecf{\beta},\tau) )\tr
   (\hat{\matf{W}}-\matf{W})    
   (\hat{\vecf{M}}(\vecf{\beta},\tau)-\vecf{M}(\vecf{\beta},\tau) )
   }\\
&= \sup_{(\beta, \tau)\in \mathcal{B}\times \mathcal{T}}  
\norm{ 
   \vecf{M}(\vecf{\beta},\tau)\tr 
   (\hat{\matf{W}}-\matf{W})    
   \vecf{M}(\vecf{\beta},\tau)
} 
+ \sup_{(\beta, \tau)\in \mathcal{B}\times \mathcal{T}} 
\norm{
  \vecf{M}(\vecf{\beta},\tau)\tr 
  \matf{W} 
  (\hat{\vecf{M}}(\vecf{\beta},\tau)-\vecf{M}(\vecf{\beta},\tau)  )
} \\
&\quad + \sup_{(\beta, \tau)\in \mathcal{B}\times \mathcal{T}} 
\norm{
  (\hat{\vecf{M}}(\vecf{\beta},\tau)-\vecf{M}(\vecf{\beta},\tau) )\tr
  \matf{W}
  \vecf{M}(\vecf{\beta},\tau) 
} 
\\&\quad + \sup_{(\beta, \tau)\in \mathcal{B}\times \mathcal{T}} 
\norm{ 
  \vecf{M}(\vecf{\beta},\tau)\tr 
  (\hat{\matf{W}}-\matf{W}) 
  (\hat{\vecf{M}}(\vecf{\beta},\tau)-\vecf{M}(\vecf{\beta},\tau) )
}\\
&\quad + \sup_{(\beta, \tau)\in \mathcal{B}\times \mathcal{T}} 
\norm{
  (\hat{\vecf{M}}(\vecf{\beta},\tau)-\vecf{M}(\vecf{\beta},\tau) )\tr
  \matf{W} 
  (\hat{\vecf{M}}(\vecf{\beta},\tau)-\vecf{M}(\vecf{\beta},\tau) )
}\\
&\quad + \sup_{(\beta, \tau)\in \mathcal{B}\times \mathcal{T}} 
\norm{
   (\hat{\vecf{M}}(\vecf{\beta},\tau)-\vecf{M}(\vecf{\beta},\tau) )\tr
   (\hat{\matf{W}}-\matf{W})    
   \vecf{M}(\vecf{\beta},\tau)
}\\
&\quad + \sup_{(\beta, \tau)\in \mathcal{B}\times \mathcal{T}} 
\norm{
   (\hat{\vecf{M}}(\vecf{\beta},\tau)-\vecf{M}(\vecf{\beta},\tau) )\tr
   (\hat{\matf{W}}-\matf{W})    
   (\hat{\vecf{M}}(\vecf{\beta},\tau)-\vecf{M}(\vecf{\beta},\tau) )
   }\\
&\le \sup_{(\beta, \tau)\in \mathcal{B}\times \mathcal{T}}  
\overbrace{
        \norm{   \vecf{M}(\vecf{\beta},\tau)\tr }
\norm{
   (\hat{\matf{W}}-\matf{W})  }
\norm{
   \vecf{M}(\vecf{\beta},\tau)
} }^{\textrm{by Cauchy--Schwarz inequality}}
+ \sup_{(\beta, \tau)\in \mathcal{B}\times \mathcal{T}} 
\norm{
  \vecf{M}(\vecf{\beta},\tau)\tr }
\norm{ \matf{W} }
\norm{ (\hat{\vecf{M}}(\vecf{\beta},\tau)-\vecf{M}(\vecf{\beta},\tau)  )
} \\
&\quad + \sup_{(\beta, \tau)\in \mathcal{B}\times \mathcal{T}} 
\norm{
  (\hat{\vecf{M}}(\vecf{\beta},\tau)-\vecf{M}(\vecf{\beta},\tau) )\tr }
\norm{  \matf{W} }
\norm{  \vecf{M}(\vecf{\beta},\tau) 
} \\
&\quad + \sup_{(\beta, \tau)\in \mathcal{B}\times \mathcal{T}} 
\norm{ 
  \vecf{M}(\vecf{\beta},\tau)\tr }
\norm{  (\hat{\matf{W}}-\matf{W}) }
\norm{  (\hat{\vecf{M}}(\vecf{\beta},\tau)-\vecf{M}(\vecf{\beta},\tau) )
}\\
&\quad + \sup_{(\beta, \tau)\in \mathcal{B}\times \mathcal{T}} 
\norm{
  (\hat{\vecf{M}}(\vecf{\beta},\tau)-\vecf{M}(\vecf{\beta},\tau) )\tr}
\norm{  \matf{W} }
\norm{  (\hat{\vecf{M}}(\vecf{\beta},\tau)-\vecf{M}(\vecf{\beta},\tau) )
}\\
&\quad + \sup_{(\beta, \tau)\in \mathcal{B}\times \mathcal{T}} 
\norm{
   (\hat{\vecf{M}}(\vecf{\beta},\tau)-\vecf{M}(\vecf{\beta},\tau) )\tr}
\norm{   (\hat{\matf{W}}-\matf{W})    }
\norm{  \vecf{M}(\vecf{\beta},\tau)
}\\
&\quad + \sup_{(\beta, \tau)\in \mathcal{B}\times \mathcal{T}} 
\norm{
   (\hat{\vecf{M}}(\vecf{\beta},\tau)-\vecf{M}(\vecf{\beta},\tau) )\tr}
\norm{   (\hat{\matf{W}}-\matf{W})  }  
\norm{   (\hat{\vecf{M}}(\vecf{\beta},\tau)-\vecf{M}(\vecf{\beta},\tau) )
   }\\
&\le \sup_{(\beta, \tau)\in \mathcal{B}\times \mathcal{T}}  
 \norm{ \vecf{M}(\vecf{\beta},\tau)\tr }
   \norm{ (\hat{\matf{W}}-\matf{W})  }
\overbrace{\sup_{(\beta, \tau)\in \mathcal{B}\times \mathcal{T}} 
   \norm{ \vecf{M}(\vecf{\beta},\tau)
} }^{=O(1)}
\\&\quad + \sup_{(\beta, \tau)\in \mathcal{B}\times \mathcal{T}} 
    \norm{ \vecf{M}(\vecf{\beta},\tau)\tr }
  \norm{ \matf{W} }
\sup_{(\beta, \tau)\in \mathcal{B}\times \mathcal{T}} 
  \norm{ (\hat{\vecf{M}}(\vecf{\beta},\tau)-\vecf{M}(\vecf{\beta},\tau)  )
} 
\\&\quad + \sup_{(\beta, \tau)\in \mathcal{B}\times \mathcal{T}} 
\norm{
  (\hat{\vecf{M}}(\vecf{\beta},\tau)-\vecf{M}(\vecf{\beta},\tau) )\tr }
 \overbrace{ \norm{  \matf{W} } }^{=O(1)}
\sup_{(\beta, \tau)\in \mathcal{B}\times \mathcal{T}}
  \norm{  \vecf{M}(\vecf{\beta},\tau) 
} \\
&\quad + \sup_{(\beta, \tau)\in \mathcal{B}\times \mathcal{T}} 
\norm{ 
  \vecf{M}(\vecf{\beta},\tau)\tr }
   \norm{  (\hat{\matf{W}}-\matf{W}) }
\sup_{(\beta, \tau)\in \mathcal{B}\times \mathcal{T}} 
   \norm{  (\hat{\vecf{M}}(\vecf{\beta},\tau)-\vecf{M}(\vecf{\beta},\tau) )
}\\
&\quad + \sup_{(\beta, \tau)\in \mathcal{B}\times \mathcal{T}} 
\norm{
  (\hat{\vecf{M}}(\vecf{\beta},\tau)-\vecf{M}(\vecf{\beta},\tau) )\tr}
   \norm{  \matf{W} }
\sup_{(\beta, \tau)\in \mathcal{B}\times \mathcal{T}} 
   \norm{  (\hat{\vecf{M}}(\vecf{\beta},\tau)-\vecf{M}(\vecf{\beta},\tau) )
}\\
&\quad + \sup_{(\beta, \tau)\in \mathcal{B}\times \mathcal{T}} 
\norm{
   (\hat{\vecf{M}}(\vecf{\beta},\tau)-\vecf{M}(\vecf{\beta},\tau) )\tr}
\norm{   (\hat{\matf{W}}-\matf{W})    }
\sup_{(\beta, \tau)\in \mathcal{B}\times \mathcal{T}} 
   \norm{  \vecf{M}(\vecf{\beta},\tau)
}\\
&\quad + \sup_{(\beta, \tau)\in \mathcal{B}\times \mathcal{T}} 
\norm{
   (\hat{\vecf{M}}(\vecf{\beta},\tau)-\vecf{M}(\vecf{\beta},\tau) )\tr}
\norm{   (\hat{\matf{W}}-\matf{W})  }  
\sup_{(\beta, \tau)\in \mathcal{B}\times \mathcal{T}} 
  \norm{   (\hat{\vecf{M}}(\vecf{\beta},\tau)-\vecf{M}(\vecf{\beta},\tau) )
   }\\
&=o_p(1)+o_p(1)+o_p(1)+o_p(1)+o_p(1)+o_p(1)+o_p(1)
 =o_p(1).
\end{align*} 
Above, we know $\norm{\matf{W}}=O(1)$ since $\matf{W}$ is fixed.
\citeposs{deCastroGalvaoKaplanLiu19} Theorem 3 proof shows that the function $\vecf{M}(\vecf{\beta},\tau)$ is continuous in $\vecf{\beta}$ given \cref{a:Y,a:Lambda}. 
Then we know $\sup_{(\beta, \tau)\in \mathcal{B}\times \mathcal{T}} 
\norm{\vecf{M}(\vecf{\beta},\tau) } = O(1)$ 
since $\vecf{M}(\vecf{\beta},\tau) $ is continuous in $(\vecf{\beta},\tau)$ and $ \mathcal{B}\times \mathcal{T}$ is a compact set.

The second condition of Theorem 5.7 in \citet{vanderVaart98} is that $(\vecf{\beta}_{0}, \tau_0) $ satisfies the well-separated minimum property. 
Since $\vecf{M}(\vecf{\beta},\tau)\tr  \matf{W}    \vecf{M}(\vecf{\beta},\tau) $ is continuous in $(\vecf{\beta},\tau)$ and 
$\{ (\vecf{\beta},\tau) : 
\norm{(\vecf{\beta}\tr,\tau)\tr-(\vecf{\beta}_{0}\tr,\tau_0)\tr} 
\ge\epsilon, 
\vecf{\beta}\in\mathcal{B}, \tau \in \mathcal{T} \}$ 
is a compact set, let $(\vecf{\beta}^{*},\tau^{*})$ denote the minimizer: for any $\epsilon>0 $,
\begin{equation}\label{eqn:Mn-pt-Thm57-cond2}
\inf_{(\vecf{\beta},\tau):
\norm{(\vecf{\beta}\tr,\tau)\tr-(\vecf{\beta}_{0}\tr,\tau_0)\tr} 
\ge\epsilon, 
}
\vecf{M}(\vecf{\beta},\tau)\tr  \matf{W}    \vecf{M}(\vecf{\beta},\tau)
= \vecf{M}(\vecf{\beta}^{*},\tau^{*})\tr  \matf{W }   \vecf{M}(\vecf{\beta}^{*},\tau^{*}) . 
\end{equation}
By \cref{a:B}, $\vecf{M}(\vecf{\beta},\tau)\ne\vecf{0}$ for any $(\vecf{\beta},\tau)\ne(\vecf{\beta}_{0},\tau_0)$, 
so $\vecf{M}(\vecf{\beta}^*,\tau^*)\ne\vecf{0}$. 
Since $\matf{W}$ is positive definite (\cref{a:W}), 
\begin{equation*} 
\vecf{M}(\vecf{\beta}^{*},\tau^*)\tr \matf{ W  }  \vecf{M}(\vecf{\beta}^{*},\tau^*) 
> 0.
\end{equation*}
Thus, for any $\epsilon>0$, 
\begin{equation}\label{eqn:Thm5.7-cond2-GMM}
\inf_{(\vecf{\beta},\tau):
\norm{(\vecf{\beta}\tr,\tau)\tr-(\vecf{\beta}_{0}\tr,\tau_0)\tr} 
\ge\epsilon
}
\vecf{M}(\vecf{\beta},\tau)\tr  \matf{W }   \vecf{M}(\vecf{\beta},\tau)
> 0
= \vecf{M}(\vecf{\beta}_{0},\tau_0)\tr \matf{ W}    \vecf{M}(\vecf{\beta}_{0},\tau_0) . 
\end{equation} 
Consistency of $\hat{\vecf{\theta}}_{\mathrm{GMM}}=(\hat{\vecf{\beta}}_{\mathrm{GMM}}\tr,\hat{\tau}_{\mathrm{GMM}})\tr$ follows by Theorem 5.7 in \citet{vanderVaart98}.
\end{proof}

\begin{proof}[Proof of \Cref{lem:Mn0-normality}]
This proof follows from Lemma 4 of \citet{deCastroGalvaoKaplanLiu19}.
Decomposing into a mean-zero term and a ``bias'' term, 
\begin{align*}
\sqrt{n}\hat{\vecf{M}}_n(\vecf{\beta}_{0},\tau_0)
= \overbrace{\sqrt{n} \bigl\{ \hat{\vecf{M}}_n(\vecf{\beta}_{0},\tau_0) 
                  -\E[ \hat{\vecf{M}}_n(\vecf{\beta}_{0},\tau_0) ] \bigr\}
            }^{\dconv \Normalp{\vecf{0}}{\matf{\Sigma}}\textrm{ by \cref{a:CLT}}}
+ \overbrace{\sqrt{n} \E[ \hat{\vecf{M}}_n(\vecf{\beta}_{0},\tau_0) ]}^{\textrm{want to show }o_p(1)} . 
\end{align*}
With iid data, we can apply \citet[Thm.\ 1]{KaplanSun17} to get  $\matf{\Sigma}=\tau_0(1-\tau_0) \E[ \vecf{Z}_i\vecf{Z}_i\tr  ]$.
The remainder of the proof shows that the second term is indeed $o_p(1)$, actually $o(1)$. 

Let $\Lambda_{ji}\equiv \Lambda(\vecf{Y}_{ji},\vecf{X}_{ji},\vecf{\beta}_{0})$, with marginal PDF $f_{\Lambda_j}(\cdot)$ and conditional PDF $f_{\Lambda_j|\vecf{Z}_j}(\cdot\mid \vecf{z})$ given $\vecf{Z}_{ji}=\vecf{z}$. 
Given strict stationarity of the data, using the definitions in \cref{eqn:def-M-hat}, assuming the support of $\Lambda_{ji}$ given $\vecf{Z}_{ji}=\vecf{z}$ is the interval $[\Lambda_{jL}(\vecf{z}),\Lambda_{jH}(\vecf{z})]$ with $\Lambda_{jL}(\vecf{z})\le-h_n\le h_n\le \Lambda_{jH}(\vecf{z})$, 
consider each component $\hat{\vecf{M}}_n^j(\vecf{\beta},\tau)$ for $j=1,2$ of the moments $\hat{\vecf{M}}_n(\vecf{\beta},\tau)=(\hat{\vecf{M}}_n^1(\vecf{\beta},\tau)\tr, \hat{\vecf{M}}_n^2(\vecf{\beta},\tau)\tr)\tr$,
\begin{align*}
& \E[ \hat{\vecf{M}}_n^j(\vecf{\beta}_{0},\tau_0) ]
= \E\left[ \frac{1}{n} \sum_{i=1}^{n} \vecf{g}_n(\vecf{Y}_{ji},\vecf{X}_{ji},\vecf{Z}_{ji},\vecf{\beta}_{0},\tau_0) \right]
= \E\left[ \vecf{g}_n(\vecf{Y}_{ji},\vecf{X}_{ji},\vecf{Z}_{ji},\vecf{\beta}_{0},\tau_0) \right] 
\\&= \E\left\{ \vecf{Z}_{ji} [ \tilde{I}(-\Lambda_{ji}/h_n) - \tau_0 ] \right\} 
\\&= \E\left\{ \vecf{Z}_{ji} \E[ \tilde{I}(-\Lambda_{ji}/h_n) - \tau_0 \mid \vecf{Z}_{ji} ] \right\} 
\\&= \E\biggl\{ \vecf{Z}_{ji} \overbrace{
               \int_{\Lambda_{jL}(\vecf{Z}_{ji})}^{\Lambda_{jH}(\vecf{Z}_{ji})} [\tilde{I}(-L/h_n) - \tau_0]
               \,dF_{\Lambda_j|Z_j}(L\mid\vecf{Z}_{ji})}^{\textrm{integrate by parts}} \biggr\} 
\\&= \E\biggl\{ \vecf{Z}_{ji} \biggl[
       \overbrace{\left. \left( \tilde{I}(-L/h_n)-\tau_0 \right) F_{\Lambda_j|Z_j}(L\mid \vecf{Z}_{ji}) 
       \right\rvert_{\Lambda_{jL}(\vecf{Z}_{ji})}^{\Lambda_{jH}(\vecf{Z}_{ji})}}^{=-\tau_0\textrm{: use \cref{a:Itilde} and $\Lambda_{jH}(\vecf{Z}_{ji})\ge h_n$}}
\\&\qquad\qquad\qquad
        -\int_{\Lambda_{jL}(\vecf{Z}_{ji})}^{\Lambda_{jH}(\vecf{Z}_{ji})}
         F_{\Lambda_j|Z_j}(L\mid \vecf{Z}_{ji}) \overbrace{\tilde{I}'(-L/h_n)}^{=0\textrm{ for }L\not\in[-h_n,h_n]} (-h_n^{-1} )
         \,dL 
        \biggr] \biggr\} 
\\&= \E\biggl\{ \vecf{Z}_{ji} \biggl[ -\tau_0
       +\overbrace{h_n^{-1} \int_{-h_n}^{h_n}
         F_{\Lambda_j|Z_j}(L\mid \vecf{Z}_{ji}) \tilde{I}'(-L/h_n) 
         \,dL }^{\textrm{change of variables to }v=-L/h_n} 
        \biggr] \biggr\} 
\\&= \E\left\{ \vecf{Z}_{ji} \left[ -\tau_0
       +\int_{-1}^{1} F_{\Lambda_j|Z_j}(-h_n v\mid \vecf{Z}_{ji}) \tilde{I}'(v) 
         \,dv \right] \right\} 
\\&= \E\left\{ \vecf{Z}_{ji} \left[ -\tau_0
       +\int_{-1}^{1} \left( \sum_{k=0}^{r}F_{\Lambda_j|Z_j}^{(k)}(0\mid \vecf{Z}_{ji})\frac{(-h_n)^k v^k}{k!} \right) \tilde{I}'(v) 
         \,dv \right] \right\} 
\\&\quad+ \E\biggl\{ \vecf{Z}_{ji} 
       \int_{-1}^{1} 
        \overbrace{f_{\Lambda_j|Z_j}^{(r)}(-\tilde{h}v \mid \vecf{Z}_{ji})}^{\tilde{h}\in[0,h_n]\textrm{ (by mean value theorem)}} 
        \frac{(-h_n)^{r+1} v^{r+1}}{(r+1)!} \tilde{I}'(v) 
         \,dv \biggr\} 
\\&= \E\biggl\{ \vecf{Z}_{ji} \biggl[ -\tau_0
       +\sum_{k=0}^{r} F_{\Lambda_j|Z_j}^{(k)}(0\mid \vecf{Z}_{ji}) \frac{(-h_n)^k}{k!}
         \overbrace{\int_{-1}^{1} v^k \tilde{I}'(v) \,dv }^{=0\textrm{ for $1\le k\le r-1$ by \cref{a:Itilde}}}
     \biggr] \biggr\} 
\\&\quad+ O(h_n^{r+1}) \overbrace{\E\biggl\{ \vecf{Z}_{ji} 
        \int_{-1}^{1} 
        \overbrace{f_{\Lambda_j|Z_j}^{(r)}(-\tilde{h}v \mid \vecf{Z}_{ji})}^{\textrm{bounded by \cref{a:U}}}
        v^{r+1} \tilde{I}'(v) 
         \,dv \biggr\} }^{O(1)\textrm{ by \cref{a:U,a:Itilde}}}
\\&= \E\left\{ \vecf{Z}_{ji} \left[ -\tau_0 +F_{\Lambda_j|Z_j}(0\mid \vecf{Z}_{ji})
       +f_{\Lambda_j|Z_j}^{(r-1)}(0\mid \vecf{Z}_{ji}) \frac{(-h_n)^r}{r!}
         \int_{-1}^{1} v^r \tilde{I}'(v) \,dv 
     \right] \right\} 
+ O(h_n^{r+1})
\\&= \E\left\{ \vecf{Z}_{ji} \left[ -\tau_0 +\E( \Ind{\Lambda_{ji}\le0} \mid \vecf{Z}_{ji} ) 
     \right] \right\} 
+ \overbrace{\frac{(-h_n)^r}{r!}}^{\textrm{$r$ is even}} \left[ \int_{-1}^{1} v^r \tilde{I}'(v) \,dv \right] 
  \E\left[ \vecf{Z}_{ji} f_{\Lambda_j|Z_j}^{(r-1)}(0\mid \vecf{Z}_{ji}) \right] 
+ O(h_n^{r+1})
\\&= \overbrace{\E\left\{ \E\left[ \vecf{Z}_{ji} \left( \Ind{\Lambda_{ji}\le0} -\tau_0 \right) \mid \vecf{Z}_{ji}  
     \right] \right\} }^{=\E\left[ \vecf{Z}_{ji} \left( \Ind{\Lambda_{ji}\le0} -\tau_0 \right) \right]=0\textrm{ by \cref{a:B}}}
+ \frac{h_n^r}{r!} \left[ \int_{-1}^{1} v^r \tilde{I}'(v) \,dv \right] 
  \E\left[ \vecf{Z}_{ji} f_{\Lambda_j|Z_j}^{(r-1)}(0\mid \vecf{Z}_{ji}) \right] 
+ O(h^{r+1})
\\&= \frac{h_n^r}{r!} \left[ \int_{-1}^{1} v^r \tilde{I}'(v) \,dv \right] 
  \E\left[ \vecf{Z}_{ji} f_{\Lambda_j|Z_j}^{(r-1)}(0\mid \vecf{Z}_{ji}) \right] 
+ O(h_n^{r+1})
= O(h_n^r)
. 
\end{align*}
Thus, the result follows if $\sqrt{n}h_n^r=o(1)$, i.e., $h_n=o(n^{-1/(2r)})$ as in \cref{a:h}. 
\end{proof}


\begin{proof}[Proof of \Cref{thm:normality}]

This proof partially follows Theorem 5 of \citet{deCastroGalvaoKaplanLiu19}.
We prove the asymptotic normality of the smoothed GMM estimator.

We start from the first-order condition for the mean value expansion. 
From the definition of $(\hat{\vecf{\beta}}_{\mathrm{GMM}},\hat{\tau}_{\mathrm{GMM}})$ in \cref{eqn:def-est-GMM}, we have the first-order condition 
\begin{equation}\label{eqn:FOC}
\left[ \vecf{\nabla}_{(\vecf{\beta}\tr,\tau)} \hat{\vecf{M}}_n(\hat{\vecf{\beta}}_{\mathrm{GMM}},\hat{\tau}_{\mathrm{GMM}}) \right]\tr \hat{\matf{W} }   \hat{\vecf{M}}_n(\hat{\vecf{\beta}}_{\mathrm{GMM}},\hat{\tau}_{\mathrm{GMM}})
= \vecf{0} . 
\end{equation}
Define 
\begin{equation}\label{eqn:nabla-Mn}
\vecf{\nabla}_{(\vecf{\beta}\tr,\tau) } \hat{\vecf{M}}_n(\vecf{\beta}_{0},\tau_0) 
\equiv
\left. \pD{}{(\vecf{\beta}\tr,\tau) } \hat{\vecf{M}}_n(\vecf{\beta},\tau)
\right\rvert_{(\vecf{\beta},\tau)=(\vecf{\beta}_{0},\tau_0)} . 
\end{equation}
Let $\hat{\vecf{M}}_n^{(k)}(\vecf{\beta},\tau)$ refer to the $k$th element in the vector $\hat{\vecf{M}}_n(\vecf{\beta},\tau)$, so $\vecf{\nabla}_{(\vecf{\beta}\tr,\tau) } \hat{\vecf{M}}_n^{(k)}(\vecf{\beta}_{0},\tau_0)$ is a row vector and $\vecf{\nabla}_{(\vecf{\beta}\tr, \tau)\tr} \hat{\vecf{M}}_n^{(k)}(\vecf{\beta}_{0},\tau_0)$ is a column vector. 
Define 
\begin{equation}\label{eqn:def-M-dot}
\dot{\matf{M}}_n(\tilde{\vecf{\beta}}_{},\tilde\tau)
\equiv \left( 
  \vecf{\nabla}_{(\vecf{\beta}\tr,\tau)\tr} \hat{\vecf{M}}_n^{(1)}(\tilde{\vecf{\beta}}_{},\tilde\tau), 
  \ldots, 
  \vecf{\nabla}_{(\vecf{\beta}\tr,\tau)\tr} \hat{\vecf{M}}_n^{(2d_Z)}(\tilde{\vecf{\beta}}_{},\tilde\tau)
\right) \tr  , 
\end{equation}
a $2d_Z\times (d_\beta+1)$ matrix with its first row equal to that of $\vecf{\nabla}_{(\vecf{\beta}\tr,\tilde\tau) } \hat{\vecf{M}}_n^{(1)}(\tilde{\vecf{\beta}}_{},\tilde\tau)$, its second row equal to that of $\vecf{\nabla}_{(\vecf{\beta}\tr,\tau) } \hat{\vecf{M}}_n^{(2)}(\tilde{\vecf{\beta}}_{},\tilde\tau)$, etc., where 
$(\tilde{\vecf{\beta}}_{},\tilde{\tau})$ lies on the line segment between 
$(\vecf{\beta}_{0}, \tau_0)$ and $(\hat{\vecf{\beta}}_{\mathrm{GMM}},\hat{\tau}_{\mathrm{GMM}})$. 
By the mean value theorem, 
\begin{equation}\label{eqn:MVT}
\hat{\vecf{M}}_n(\hat{\vecf{\beta}}_{\mathrm{GMM}},\hat\tau_{\mathrm{GMM}})=
\hat{\vecf{M}}_n(\vecf{\beta}_{0},\tau_0)
+\dot{\matf{M}}_n
(\tilde{\vecf{\beta}},\tilde\tau)
\begin{bmatrix}
 \hat{\vecf{\beta}}_{\mathrm{GMM}} - \vecf{\beta}_{0} \\
 \hat{\tau}_{\mathrm{GMM}} - \tau_{0}
\end{bmatrix}.
\end{equation}
Pre-multiplying \cref{eqn:MVT} by $[ \vecf{\nabla}_{(\vecf{\beta}\tr,\tau)} \hat{\vecf{M}}_n(\hat{\vecf{\beta}}_{\mathrm{GMM}},\hat{\tau}_{\mathrm{GMM}}) ]\tr \hat{\matf{W} }$ and using \cref{eqn:FOC} for the first equality, 
\begin{align*}
\vecf{0} 
&=[ \vecf{\nabla}_{(\vecf{\beta}\tr,\tau)} \hat{\vecf{M}}_n(\hat{\vecf{\beta}}_{\mathrm{GMM}},\hat{\tau}_{\mathrm{GMM}}) ]\tr \hat{\matf{W} }   \hat{\vecf{M}}_n(\hat{\vecf{\beta}}_{\mathrm{GMM}},\hat{\tau}_{\mathrm{GMM}})\\
&= [ \vecf{\nabla}_{(\vecf{\beta}\tr,\tau)} \hat{\vecf{M}}_n(\hat{\vecf{\beta}}_{\mathrm{GMM}},\hat{\tau}_{\mathrm{GMM}}) ]\tr \hat{\matf{W} }
\hat{\vecf{M}}_n(\vecf{\beta}_{0},\tau_0)
+[ \vecf{\nabla}_{(\vecf{\beta}\tr,\tau)} \hat{\vecf{M}}_n(\hat{\vecf{\beta}}_{\mathrm{GMM}},\hat{\tau}_{\mathrm{GMM}}) ]\tr \hat{\matf{W} }
\dot{\matf{M}}_n(\tilde{\vecf{\beta}},\tilde\tau) ( \hat{\vecf{\theta}}_{\mathrm{GMM}} - \vecf{\theta}_{0} ).
\end{align*}
Multiplying by $\sqrt{n}$ and rearranging, 
\begin{align}\notag
& \sqrt{n} ( \hat{\vecf{\theta}}_{\mathrm{GMM}} - \vecf{\theta}_{0} )
\\[-12pt]&= -\bigl\{ [ \vecf{\nabla}_{(\vecf{\beta}\tr,\tau)} \hat{\vecf{M}}_n(\hat{\vecf{\beta}}_{\mathrm{GMM}},\hat{\tau}_{\mathrm{GMM}}) ]\tr \hat{\matf{W} }
\dot{\matf{M}}_n(\tilde{\vecf{\beta}},\tilde\tau) \bigr\}^{-1}
[ \vecf{\nabla}_{(\vecf{\beta}\tr,\tau)} \hat{\vecf{M}}_n(\hat{\vecf{\beta}}_{\mathrm{GMM}},\hat{\tau}_{\mathrm{GMM}}) ]\tr \hat{\matf{W} }
\overbrace{\sqrt{n} \hat{\vecf{M}}_n(\vecf{\beta}_{0},\tau_0)}^{=O_p(1)}
\label{eqn:est-GMM-asy-linear-almost}
\\&= -\{ \matf{G}\tr \matf{W} \matf{G} \}^{-1}
\matf{G}\tr \matf{W}
\sqrt{n} \hat{\vecf{M}}_n(\vecf{\beta}_{0},\tau_0)
+o_p(1) ,
\label{eqn:est-GMM-asy-linear}
\end{align}
where $\hat{\matf{W}} = \matf{W} +o_p(1)$ by \cref{a:W}, 
and $\dot{\matf{M}}_n(\tilde{\vecf{\beta}},\tilde\tau) = \matf{G} +o_p(1)$ and 
$\vecf{\nabla}_{(\vecf{\beta}\tr,\tau)} \hat{\vecf{M}}_n(\hat{\vecf{\beta}}_{\mathrm{GMM}},\hat{\tau}_{\mathrm{GMM}}) 
= \matf{G} +o_p(1)$ 
by \cref{a:G-est}. 
From \cref{lem:Mn0-normality}, 
$\sqrt{n}\hat{\vecf{M}}_n(\vecf{\beta}_{0},\tau_0)
\dconv 
\Normalp{\vecf{0}}{\matf{\Sigma}}$. 
Applying the continuous mapping theorem yields 
the stated result. 
\end{proof}

\section{Sensitivity to bandwidth}
\label{sec:bandwidth}

This section presents the simulation and empirical results with various bandwidth.
For the simulation, we consider the fixed (small) smoothing bandwidth $h=\{0.1, 1\}$ and the plug-in bandwidth from \cite{KaplanSun17}.
For the empirical example, we consider the set of fixed bandwidth $h=\{0.01, 0.1\}$ and the plug-in bandwidth. 
We report the minimum, maximum, and mean value of the plug-in bandwidth.

\citet{KaplanSun17} has shown that their plug-in bandwidth is MSE-optimal for estimation in their setting.
We use \citeposs{KaplanSun17} plug-in bandwidth as a suggestive upper bound for our fixed bandwidth. 
\citet{KaplanSun17} also note that using a small smoothing bandwidth yields estimates similar to those of the unsmoothed estimator in \cite{ChernozhukovHansen06}.
Compared to the unsmoothed method, one of the main advantages of the smoothing method is to reduce the computation time, making simulations with a large number of replications feasible.

\cref{tab:sim:DGP:all,tab:emp:bs:allh} illustrate that the simulation and empirical example estimation results are not very sensitive to the smoothing bandwidth used, especially when there is little smoothing.

\begin{table}[!htbp] 
    \centering
\caption{\label{tab:sim:DGP:all}  Percentage bias and RMSE of parameters in simulation.}
\begin{threeparttable}
\begin{tabular}[c]{llS[table-format=4.0,round-precision=0] 
  cS[table-format=3.1,round-precision=1]S[table-format=1.3,round-precision=3]
  cS[table-format=3.1,round-precision=1]S[table-format=1.3,round-precision=3]
  cS[table-format=3.1,round-precision=1]S[table-format=1.3,round-precision=3]
  }
  \toprule
&   &  &&  \multicolumn{2}{c}{$\hat{\tau}$}          && \multicolumn{2}{c}{$\hat{\beta}_1$} && \multicolumn{2}{c}{$\hat{\beta}_0$} \\
  \cmidrule{5-6} \cmidrule{8-9} \cmidrule{11-12}
{DGP} & {$h$} & \multicolumn{1}{c}{$n$} &&  {\% Bias} & {RMSE} &&  {\% Bias} & {RMSE} && {\% Bias} & {RMSE}  \\
  \midrule
1 & 0.1  & 1500        &&    0.023429  &    0.037891  &&    0.048429  &    0.029260  &&    0.457093  &    0.158950     \\
  &      & 3000        &&    0.093429  &    0.026950  &&   -0.061714  &    0.020629  &&    1.061402  &    0.114492     \\
  &      & 5000        &&   -0.041857  &    0.020240  &&   -0.022429  &    0.015820  &&    0.053585  &    0.086821     \\[0.5em]
  & 1    & 1500        &&    0.677714  &    0.041725  &&    0.241571  &    0.028186  &&    2.809494  &    0.158110     \\
  &      & 3000        &&    0.247000  &    0.025760  &&   -0.007571  &    0.019223  &&    1.638061  &    0.108769     \\
  &      & 5000        &&    0.114000  &    0.019417  &&    0.032571  &    0.014965  &&    0.648741  &    0.082216     \\[0.5em]
  & KS17 & 1500        &&    1.864429  &    0.043310  &&    0.815714  &    0.027974  &&    5.214716  &    0.153422     \\
  &      & 3000        &&    0.651714  &    0.026052  &&    0.229571  &    0.018712  &&    2.562927  &    0.105266     \\
  &      & 5000        &&    0.385143  &    0.019528  &&    0.185429  &    0.014722  &&    1.326276  &    0.080348     \\

  \midrule
2 & 0.1  & 1500        &&    2.830571  &    0.121328  &&    3.216600  &    0.701847  &&    2.604457  &    1.240732     \\
  &      & 3000        &&  -14.412000  &    0.186718  &&  -16.021914  &    1.043603  &&  -14.056829  &    1.232065     \\
  &      & 5000        &&  -27.681571  &    0.215062  &&  -30.583857  &    1.196884  &&  -28.250200  &    1.278548     \\[0.5em]
  & 1    & 1500        &&   -2.016571  &    0.099206  &&   -1.987771  &    0.582542  &&   -2.980714  &    1.214856     \\
  &      & 3000        &&   -4.857571  &    0.105214  &&   -5.553143  &    0.596574  &&   -4.070171  &    0.962094     \\
  &      & 5000        &&   -4.443571  &    0.074541  &&   -4.880000  &    0.427539  &&   -4.670057  &    0.729385     \\[0.5em]
  & KS17 & 1500        &&   -1.022857  &    0.088957  &&   -0.746143  &    0.533015  &&   -4.100143  &    1.022516     \\
  &      & 3000        &&   -0.971857  &    0.063449  &&   -0.745343  &    0.380732  &&   -3.214171  &    0.766956     \\
  &      & 5000        &&   -0.798143  &    0.048720  &&   -0.497829  &    0.295990  &&   -2.990457  &    0.598567     \\

\bottomrule
\end{tabular}
\begin{tablenotes}
\item 1000 replications. The plug-in bandwidth from \citet{KaplanSun17} ranges around 0.92 to 2.61 with a mean value of around 1.48 in DGP 1 and ranges around 6.0 to 37.7 with a mean value of 10.8 in DGP 2.
Percentage bias is computed as the bias divided by the true parameter, then multiplied by 100. 
\end{tablenotes}
\end{threeparttable}
\end{table}

\begin{table}[!htbp] 
    \centering
\caption{\label{tab:emp:bs:allh} Estimates of parameters in empirical example.}
\begin{threeparttable}
\begin{tabular}[c]{lS[table-format=4.0,round-precision=0] 
    cS[table-format=1.3,round-precision=3]S[table-format=1.3,round-precision=3]
    S[table-format=1.3,round-precision=3]
    >{{[}} 
    S[table-format=4.3,round-precision=3]
    @{,\,} 
    S[table-format=2.3,round-precision=3]
    <{{]}} 
  }
  \toprule
\multicolumn{1}{c}{Bandwidth}   &     \multicolumn{1}{c}{Parameters}          && {Estimates} & {BS.SE} && \multicolumn{2}{c}{CI} \\
  \midrule
$h=0.01$  &$\tau$           &&    0.410726  &    0.039481  &&    0.333343  &    0.488109     \\
          &$\delta$         &&    1.175411  &    0.178195  &&    0.826150  &    1.524673     \\
          &$\gamma$         &&    1.275897  &    0.967189  &&   -0.619793  &    3.171587     \\
          &{EIS}            &&    0.783762  &    0.385821  &&    0.027553  &    1.539972     \\

\midrule
$h=0.1$   &$\tau$           &&    0.414411  &    0.031076  &&    0.353502  &    0.475321     \\
          &$\delta$         &&    1.189850  &    0.075203  &&    1.042452  &    1.337248     \\
          &$\gamma$         &&    1.446061  &    0.468206  &&    0.528377  &    2.363744     \\
          &{EIS}            &&    0.691534  &    0.247131  &&    0.207156  &    1.175912     \\
\midrule
KS17      &$\tau$           &&    0.402151  &    0.030903  &&    0.341581  &    0.462721     \\
          &$\delta$         &&    1.197456  &    0.075172  &&    1.050118  &    1.344794     \\
          &$\gamma$         &&    1.205733  &    0.473464  &&    0.277744  &    2.133722     \\
          &{EIS}            &&    0.829371  &    0.248475  &&    0.342360  &    1.316382     \\

\bottomrule
\end{tabular}
\begin{tablenotes}
\item 1000 bootstraps to obtain the bootstrapped standard error. Use estimates $\pm$ 1.96 times bootstrapped standard error to obtain 95\% CI. The \citeposs{KaplanSun17} plug-in bandwidth ranges from 0.4 to 1.6.
\end{tablenotes}
\end{threeparttable}
\end{table}

\section{Additional details for Monte Carlo}\label{sec:MCgraph}

\begin{figure}[!htbp]
\centering
\includegraphics[width=0.45\textwidth,height=0.3\textheight]{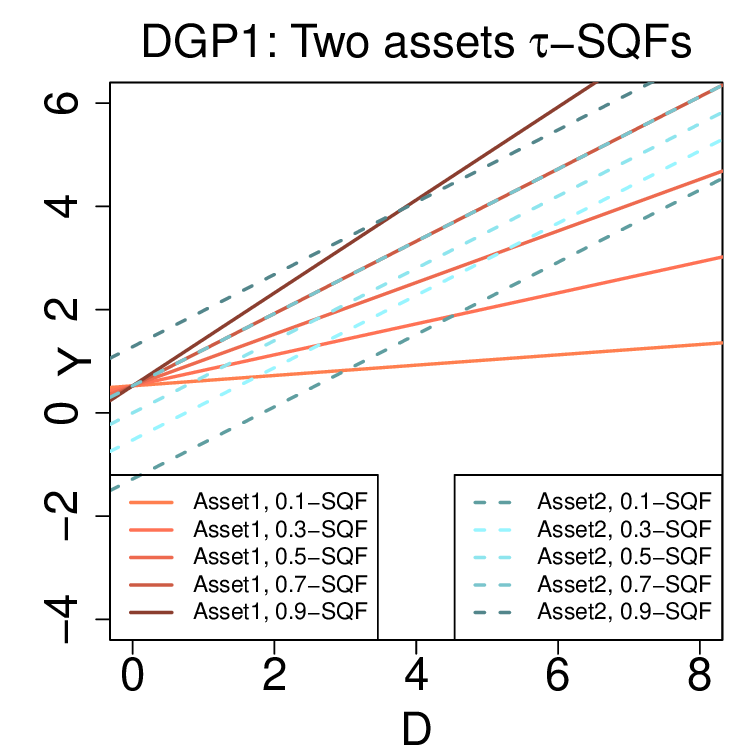}
\hfill
\includegraphics[width=0.45\textwidth,height=0.3\textheight]{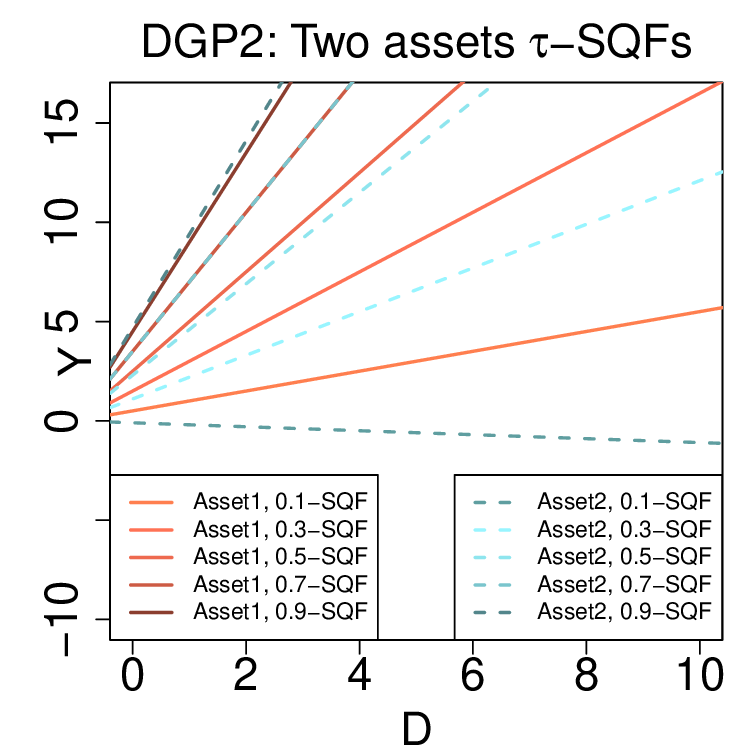}\\
\includegraphics[width=0.45\textwidth,height=0.3\textheight]{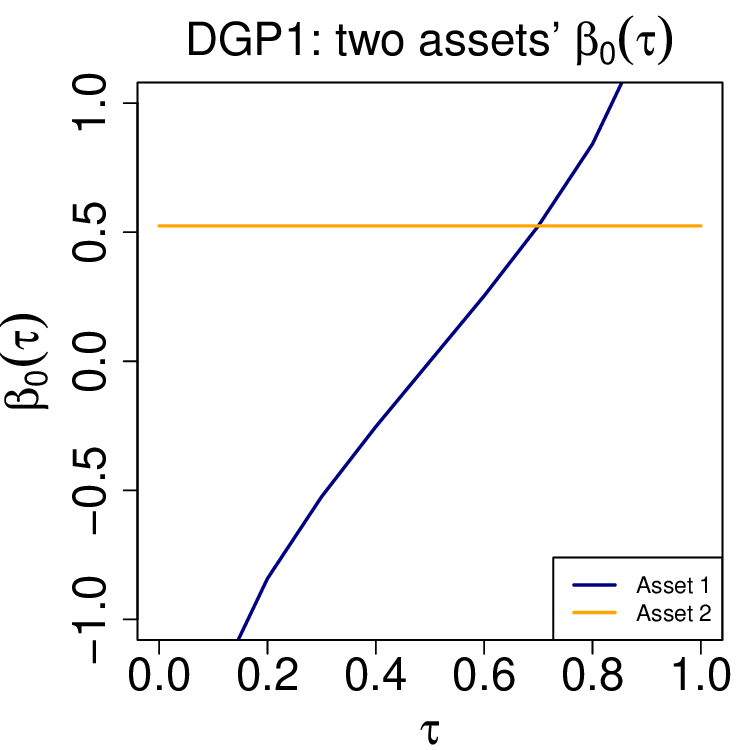}
\hfill
\includegraphics[width=0.45\textwidth,height=0.3\textheight]{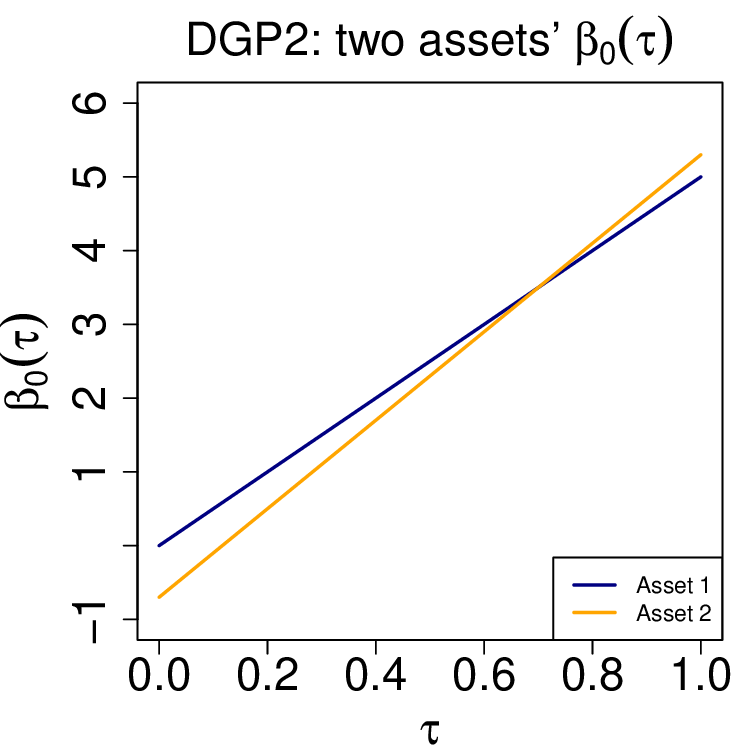}\\
\includegraphics[width=0.45\textwidth,height=0.3\textheight]{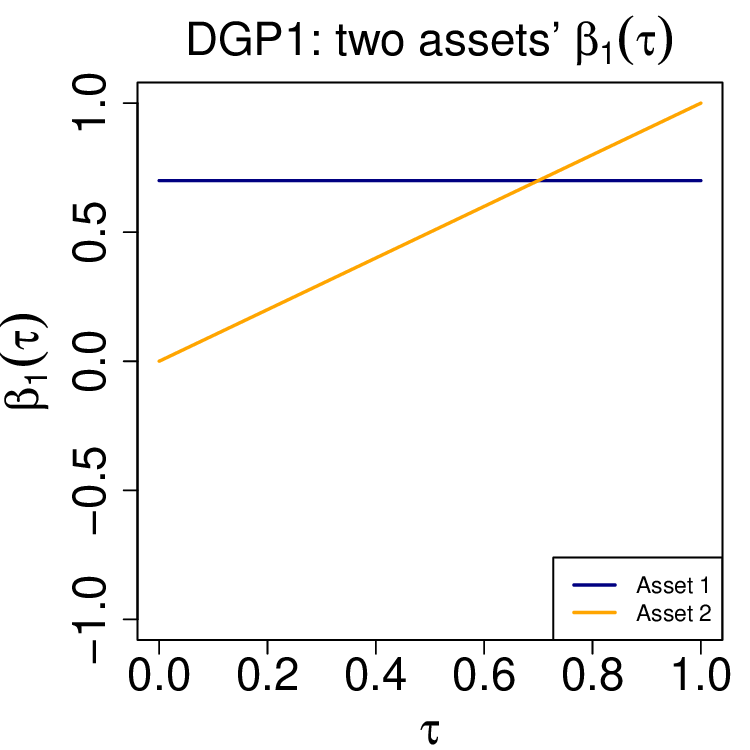}
\hfill
\includegraphics[width=0.45\textwidth,height=0.3\textheight]{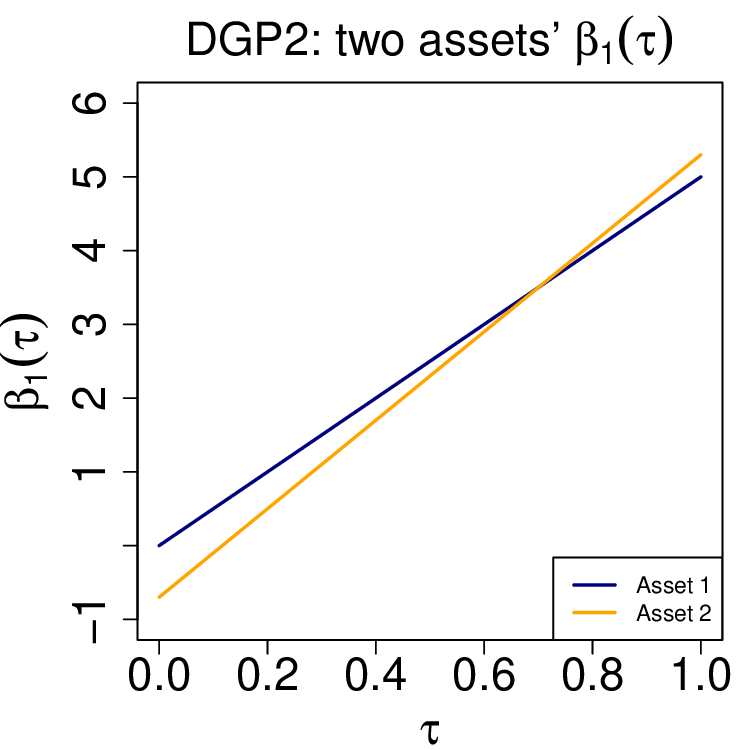}
\caption{The two assets' SQFs and coefficient functions in simulation DGP 1 and DGP 2.\label{fig:sim:all}}
\end{figure}

\cref{fig:sim:all} presents the plot of various $\tau$-structural quantile function and the coefficients as a function of $\tau$ for both assets in both DGPs.

\renewcommand{\thetable}{\arabic{table}}
\renewcommand{\thefigure}{\arabic{figure}}



\section{Introduction to  Quantile Preferences}\label{app:preferences}

This section reviews some features of the economic quantile model, see \citet{deCastroGalvaoNunes25} for further details. We start by reviewing simple properties of quantiles, and then briefly discuss the static quantile preferences. Finally, we review the notion of risk for this preference. 

\subsection{Properties of Quantiles}

We begin by stating some preliminary definitions of conditional quantiles. 
Given two correlated random variables, $W$ and $Z$, let $F(w|Z=z) = F_{W|Z=z}(w) = \Pr \left( W \leqslant w | Z=z \right)$ denote the  conditional cumulative distribution function (c.d.f.) of $W$ given $Z$. 
If the function $w \mapsto F_{W|Z=z} (w)$ is strictly increasing and continuous in its support, its inverse is the quantile of $W$ given $Z$, that is, 
\begin{equation*}
\Q[W|Z=z] = F_{W|Z=z}^{-1} (\tau),
\end{equation*}
for $\tau \in (0,1)$.\footnote{In this paper we will not consider the cases $\tau =0$ or $\tau =1$.}
This case is illustrated in Figure \ref{cdf_quantile_not_invertible}(a). 
For simplicity, in the rest of the paper we will denote $\Q[W|Z=z] $ by $\Q[W|z]$. 
If $w \mapsto F_{W|Z=z} (w)$ is not invertible, 
we can still define the quantile as one of its  generalized inverses. 
Following the standard practice, we define the quantile as the left-continuous version of the generalized inverse, that is, for $\tau \in (0,1)$ we define
\begin{equation*}
    \Q [W | Z=z] \equiv \inf \{w \in \R : \Pr[W  \leqslant w | Z= z] \geqslant \tau\}. 
\end{equation*}
Figure \ref{cdf_quantile_not_invertible}(b) illustrates the case in which $F$ is not invertible.

\begin{figure}[H]
\begin{center}

\begin{tikzpicture}[scale=0.7]

\draw[->] (-0.1,0) node[left]{$0$}-- (3.2,0) node[below]{$w$};

\draw[->] (0, 0)  -- (0,3) node[right]{$F $};



\draw (0.5,0) .. controls (0.8,0) and (1,0.9) .. (1.5,1) .. controls (2,1.1) and (2,1.9) .. (2.5,2); 


 \draw[very thick] (0.5,-0.01) -- (2.5,-0.01);
 \draw (0.5,0) node[below]{$x_0$};
 \draw[dashed] (2.5,0) node[below]{$x_1$} -- (2.5,2) -- (0,2) node[left]{$1$};

\begin{scope}[xshift=4.5cm,rotate=90,yscale=-1]
    
 \draw (0.5,0) .. controls (0.8,0) and (1,0.9) .. (1.5,1) .. controls (2,1.1) and (2,1.9) .. (2.5,2); 

\end{scope}

\begin{scope}[xshift=4.5cm]

\draw (0,0.5) node[left]{$x_0$};
\draw[dashed] (0, 2.5) node[left]{$x_1$}--(2,2.5) -- (2,0) node[below]{$1$};
\draw[->] (-0.1,0) -- (3.2,0) node[below]{$\tau$};

\draw[->] (0,-0.1) node[below]{$0$} -- (0,3) node[right]{$Q$} ;

\draw[very thick] (0,0.5)--(0,2.5);
\end{scope}

\draw (4, -1.3) node{(a)};

\begin{scope}[xshift=10cm]

\draw[->] (-0.1,0) -- (4.7,0) node[below]{$w$};

\draw[->] (0, 0) node[below]{$ 1$} -- (0,3) node[right]{$F $};

\draw[thick,black] (0,0) -- (1,0.5) -- (2,0.5);

\draw[thick,black] (2,1.5) -- (3,1.5) -- (4,2);

\draw[dashed] (0,0.5) node[left]{$0.25$} -- (1,0.5);
\draw[dashed] (0,1.5) node[left]{$0.75$} -- (2, 1.5);
\draw[dashed] (0, 2)   node[left]{$1$} -- (4,2);

\draw[dashed] (1,0) node[below]{$2 $} -- (1,0.5);
\draw[dashed] (2,0) node[below]{$3 $} -- (2,1.5);
\draw[dashed] (3,0) node[below]{$4 $} -- (3,1.5);
\draw[dashed] (4,0) node[below]{$5 $} -- (4,2);

\filldraw[black] (2,1.5) circle (0.5mm);
\filldraw[white] (2,0.5) circle (0.5mm);
\draw[black] (2,0.5) circle (0.5mm);

\draw[->] (5.9,0) -- (10.5,0) node[below]{$\tau$};

\draw[->] (6,-0.1) -- (6,3) node[right]{$Q$} ;

\draw (6,0) node[left]{$1$};

\draw[thick,black] (6,0)  -- (7,0.5);
\draw[thick,black] (7,1) -- (9,1);
\draw[thick,black] (9,1.5) -- (10,2);

\draw[dashed] (6,0.5) node[left]{$2$} -- (7,0.5);
\draw[dashed] (6,1) node[left]{$3$} -- (7,1);
\draw[dashed] (6,1.5) node[left]{$4$} -- (9,1.5);
\draw[dashed] (6,2) node[left]{$5$} -- (10,2);

\draw[dashed] (7,0) node[below]{$0.25$} -- (7,1);
\draw[dashed] (9,0) node[below]{$0.75$} -- (9,1.5);
\draw[dashed] (10,0) node[below]{$1$} -- (10,2);

\filldraw[black] (7,0.5) circle (0.5mm);
\filldraw[white] (7,1) circle (0.5mm);
\draw[black] (7,1) circle (0.5mm);

\filldraw[black] (9,1) circle (0.5mm);
\filldraw[white] (9,1.5) circle (0.5mm);
\draw[black] (9,1.5) circle (0.5mm);

\draw (5.5, -1.3) node{(b)}; 

 \end{scope}

\end{tikzpicture}

\end{center}
\caption{c.d.f. ($F$) and quantile ($Q$) functions when (a) $F$ is continuous and strictly increasing in its support $[x_0,x_1]$; and (b) $F$ is not invertible.}
\label{cdf_quantile_not_invertible}
\end{figure}

An important property of quantiles that will be repeatedly used below is the following: for any non-decreasing and left-continuous function $f: \R \to \R$ and correlated $W$ and $Z$, we have:\footnote{See \citet[Lemma A.2, p. 1927]{deCastroGalvao19}.}
\begin{align}\label{eq:prop_Qf=fQ}
    \Q \Big[ f( W) \; \big| \; Z=z \Big]  =f\left( \Q \big[W \; \big| \; Z=z \big] \right).
\end{align}
Notice that an analogous property holds for expectation only if $f$ is affine.\footnote{
This invariance with respect to monotonic continuous transformations is a useful property that quantiles have.
On the other hand, quantile fail to satisfy other interesting properties, such as additivity and the law of iterated expectation; see \citet{deCastroGalvao19}  and \citet{deCastroCostaGalvaoZubelli22} for discussion and further results.  
}

\subsection{Quantile Preferences}

The expected utility (EU) is a widely used preference in economics and econometrics. In this paper, we focus on the quantile preferences (QP). 
To contextualize the QP, let $\mathbb{R}$ denote the set of random variables (lotteries). The QP, in the simple static case, can be represented as
\begin{equation}\label{eq:quantile preference}
X \succeq Y \iff \Q[U(X)] \geqslant \Q[U(Y)].
\end{equation}
This preference was axiomatized by \citet{Chambers:09}, \citet{Rostek:10}, and \citet{deCastroGalvao22}. The latter work axiomatizes both the static and dynamic quantile preferences.

Notice that \eqref{eq:prop_Qf=fQ} implies that the use of a utility function in equation \eqref{eq:quantile preference} leads to the same preference order, as long as $u$ is increasing and continuous. Thus, a static quantile preference between random variables is defined as follows:
\begin{align*}
    X \pref Y \iff \Q[X] \geqslant \Q[Y].
\end{align*}

The QP is different from the usual EU. In the static case, an important distinction is the notion of risk. While the later depends on the curvature of the utility function, the former depends on the $\tau$-quantile. Although the expectation is simple to compute and a parsimonious statistic, QP has several advantages as: robustness, ability to separate tastes from beliefs, ability to deal with categorical (instead of continuous) variables, and the flexibility of offering a family of preferences indexed by quantiles. Next, we discuss the notion of risk associated with the QP.

\subsection{Downside Risk Attitude}\label{sec:risk in dynamic}

We review the notion of risk attitude in the quantile model. Risk attitudes in the static quantile model were first studied by \citet{Mendelson:87}, \citet{Manski88}, and \citet{Rostek:10}. To cover these aspects, and for completeness, we revisit some of the discussion contained in \citet{deCastroGalvao22} and \citet{deCastroGalvaoOta24} to provide an overview of the risk attitude in the quantile model.

\citet{Manski88} was the first to study quantile preferences and observed that the risk attitude for these preferences is determined by the quantile $\tau$ that is maximized. We will explain this in two steps: first, we show that the concavity of the utility function has no relation with risk aversion and, second, that $\tau$ captures downside risk aversion.

To see that a utility function's curvature does not have a relationship with the risk attitude for quantile preferences, it is sufficient to recall the property that quantiles commute with monotonic and continuous functions in \eqref{eq:prop_Qf=fQ}. 
Let us assume that $U:\mathbb{R} \to \mathbb{R}$ is strictly increasing and continuous, and the same applies to its inverse. 
Then, we can use the invariance with respect to the monotone transformation property of quantiles, as discussed above, to obtain:
\begin{align*}
\Q [U(X)] \geqslant \Q [U(Y)] 
\Leftrightarrow 
U^{-1} ( \Q [U(X)] )  \geqslant 
U^{-1} (
\Q [U(Y)] )  
\Leftrightarrow 
\Q [X] \geqslant \Q [Y] .
\end{align*}
Therefore, the utility function is irrelevant to the risk attitude in quantile preference. 
For a more extensive discussion on this topic, see \citet{deCastroGalvaoNunes25}.

To relate $\tau$ to downside risk aversion, we consider  
\citet[p. 354]{Rostek:10}'s definition of \emph{downside risk} or the equivalent definition of  \emph{quantile-preserving spreads} introduced by \citet{Mendelson:87}. 

\begin{definition}[Quantile-preserving spread / More downside risk]
\label{def:quantile_preserving_spread}
 The random variable   $Y$ is a \emph{quantile preserving spread} of $X$ \citep{Mendelson:87} or 
 $Y$ \emph{involves more downside risk than} $X$ with respect to $q$ \citep{Rostek:10}
 if: $(i)$ $\Q[X]=\Q[Y]=q$; $(ii)$  $F_X (x) \leqslant F_Y(x)$ for any  $x \leqslant q$ and  $(iii)$  $F_X (x) \geqslant F_Y(x)$ for any  $x \geqslant q$. 
\end{definition}

\begin{figure}[H]
    \centering

 \begin{tikzpicture}[scale=0.85]
 
 \draw[->]  (0,-0.1)node[below]{$0$} -- (0,2.5);
 
 \draw[->]  (-2,0) -- (7.5,0);
 
 \draw[dashed] (-2,2) -- (7.3,2);
 \draw (-0.2,2) node[above]{$1$};
 
 \draw[very thick,black] (-2,0) .. controls (4,2) and (3,0) .. (7,2) node[below]{$F_{Y}$};  
 
 \draw[densely dashed,color=black,very thick] (0,0) .. controls (5,0.5) and (4,0) .. (5,2) ;

 \draw[very thick, black]  (4.9,1.7) node[left]{$F_{X}$};

 \draw[dotted] (4.6,0) node[below]{$q$}
 -- (4.6,1.12) -- (0,1.12) node[left]{$\tau$};
 \draw[black] (4.6,1.1) circle (0.5mm);
 \filldraw[black] (4.6,1.1) circle (0.5mm);
 
 \draw[dotted] (-0.87,0) -- (-0.87,0.3) -- (3.5,0.3) -- (3.5,0);
 \draw[dotted] (0,1.5) -- (6,1.5) -- (6,0);
 
 \draw (0,1.5) node[left]{\scriptsize{$\tau'$}};
 \draw (0,0.3) node[left]{\scriptsize{$\tau$''}};
 
 \draw (-0.87,0) node[below]{\scriptsize{$\Q[Y]$}};
 \draw (3.5,0) node[below]{\scriptsize $\Q[X]$};

 \end{tikzpicture}

    \caption{$Y$ is a $\tau$-quantile preserving spread of $X$}
    \label{fig:tau_quantile_spread}
\end{figure}

Figure \ref{fig:tau_quantile_spread} illustrates the concept. 
In this picture, we can see that the c.d.f. of $X$, $F_X$, crosses the c.d.f. of $Y$, $F_Y$, from below exactly at $q=\Q[X]=\Q[Y]$.
Thus, $Y$ is more likely to lead to worse outcomes below $q$ and higher outcomes above $q$. That is, $Y$ is more widespread than $X$.
This justifies the notion that $Y$ is a quantile- preserving of $X$ (since the $\tau$-quantile is preserved). 
Also, $Y$ involves more downside risk than $X$ because it leads to worse outcomes with higher probability.

It is useful to see that a decision maker (DM) who maximizes the $\tau'$-quantile for a $\tau'<\tau$  prefers the less risky $X$ to the riskier $Y$. 
On the other hand, 
a DM that maximizes a $\tau''$-quantile for $\tau''>\tau$, prefers the riskier $Y$. 
That is, a DM maximizing a lower quantile prefers the less risky option; a DM maximizing a higher quantile prefers a riskier option. 
This clarifies the sense in which $\tau$ captures the risk attitude of the DM.
This result is summarized in the following:\footnote{The proof of Lemma \ref{lemma:sufficient_downside_risk_aversion} and all other results in the paper are compiled in the appendix.}

\begin{lemma}[de Castro and Galvao (2022)]\label{lemma:sufficient_downside_risk_aversion}
    If $Y$ has higher downside risk than $X$, then a quantile maximizer DM that has sufficiently high downside risk aversion, that is, a sufficiently low $\tau$, prefers $X$ to $Y$. 
\end{lemma}

In fact, 
we can relate more downside risk aversion to quantiles in slightly different forms. 
Following \citet{Rostek:10}, we can say define the following:

\begin{definition}
    A preference $\pref^{1}$ \emph{has more downside risk aversion}  than preference $\pref^{2}$ if for all 
pair of random variables $X$ and $Y$ such that $Y$ involves more downside risk than $X$, if $X \pref^{2} Y$ then 
$X \pref^{1} Y$.
\end{definition}

The intuition for this definition should be clear: whenever the DM with preference $\pref^2$ prefers the safer $X$, then so should a DM with preference $\pref^1$. 
With this definition in place, we have  the following:

\begin{lemma}[de Castro and Galvao (2022)]\label{lemma:more downside risk aversion}
   Let $\pref^1$ and $\pref^2$ two quantile preferences, maximizing  $\tau^1$ and $\tau^2$, respectively. 
   Then $\pref^{1}$  has more downside risk aversion   than preference $\pref^{2}$ if and only if $\tau^1 \leqslant \tau^2$. 
\end{lemma}

It is usual to refer to the risk attitude of quantile preferences as downside risk aversion to recognize how different it is from the standard risk aversion in expected utility models.

\subsection{Separation of Risk and Elasticity of Intertemporal Substitution}\label{sec:separation}

Since the seminal works of \citet{KrepsPorteus:78} and \citet{EpsteinZin89}, it is well understood that a separation between risk and intertemporal attitudes is only possible if the uncertainty resolution timing matters. More recently,  \citet[Proposition 3]{BommierKochovLeGrand17}  show that scale-invariant certainty equivalents generate what they call a restricted ``indifference toward the timing of resolution of uncertainty'' (ITTRU), from. This is, in a sense, the weakest form of indifference that still accommodates the separation between risk and intertemporal substitution attitudes. It is important to highlight that since the quantile certainty equivalent is scale-invariant, it belongs to this selected class, thus allowing for the separation of EIS and risk attitude.

Hereby we briefly illustrate how the aforementioned separation can be achieved. Consider the utility index $U(C) = C^{\rho}$. When $\rho \in (0,1)$, we have the case of risk aversion in the standard expected utility model. In particular, when $\rho^{1}< \rho^{2}$, individual 1 is more risk averse than 2, in the sense that individual 1 has a higher coefficient of relative risk aversion. 
However, in the static quantile preferences, any $\rho>0$ leads to exactly the same choices, as discussed above; see \citet*{deCastroGalvao22}. 
In other words, the parameter $\rho$ does not capture any aspect of the decision-maker's attitude towards risk for the static quantile preferences.\footnote{As discussed in Section \ref{sec:risk in dynamic}, the attitude towards risk is captured by $\tau$ in the quantile model.} 

Whenever multiple periods are considered, the parameter $\rho$ plays an important role. We shall illustrate this now. 
For simplicity, let us focus on the case of only two periods -- the case of multiple or even infinite periods can be dealt with in a similar fashion -- and no labor.
Consider the following quantile recursive equation, 
\begin{equation*}
V(C_{0}, \tilde{C}_{1}) =     C_{0}^{\rho} + \delta \Q \left[ \tilde{C}_{1}^{\rho} \right] .
\end{equation*}
Applied to a deterministic prospect, that is, $\tilde{C}_{1}=C_{1}$, the equation above yields $ V(C_{0}, \tilde{C}_{1}) =c_{0}^{\rho} + \delta   C_{1}^{\rho}$, an intertemporal isoelastic utility function. Recall that EIS measures the elasticity of the ratio $(C_{1}/C_{0})$ to a change in the marginal rate of substitution between $C_{0}$ and $C_{1}$, that is $MRS_{C_{1},C_{0}}= V'(C_{1})/V'(C_{0})$. Therefore, using the standard definition, $\text{EIS}=-d\ln({C_{1}/C_{0}})/d\ln MRS_{C_{1},C_{0}}$. It is easy to see that the EIS in this case is simply $\frac{1}{1-\rho}$. The EIS measures how strongly individuals are willing to intertemporally substitute consumption between two periods, the current one and the next one.\footnote{Another interpretation of the EIS is that it measures the sensitivity of consumption growth to changes in the interest rate (the return of investment opportunities).} 

It is useful to compare this quantile method  with the most widely used method to separate risk aversion and the EIS, which  is  the following specification of  \citet{EpsteinZin89} and \citet{Weil:90}, with $\rho \not =0$, $\alpha \not= 0$:
\begin{equation*}
V^{EZ} (C_{0},\tilde{C}_{1}) = \left( C_{0}^{\rho} + \delta \left( \E \left[ \tilde{C}_{1}^{\alpha}\right] \right)^{ \frac{\rho}{\alpha}} \right)^{\frac{1}{\rho}}, 
\end{equation*}
where the parameter $\rho$ determines the EIS, $1/(1-\rho)$, and the parameter $\alpha$ captures the risk aversion; with smaller values of $\alpha$, other things equal, implying a stronger aversion to risk.

As observed by \citet{BommierKochovLeGrand17}, this model satisfies monotonicity if and only if $\rho=\alpha$, in which case the model collapses to the standard expected utility model. It is well known, however, that the separation of risk aversion and EIS is not possible in the standard expected utility model. In other words, for achieving its separation goal, Esptein-Zin-Weil preferences are necessarily non-monotonic. \citet[Lemmas 2 and 3]{BommierKochovLeGrand17} show some of the problems that arise from this lack of monotonicity. In short, a Epstein-Zin-Weil decision maker may prefer to reduce lifetime utility in every state of the world just out of the willingness to reduce risk. This appears to be a shortcoming of Epstein-Zin preferences. In contrast, the willingness to reduce risk by a decision maker with monotonic preferences will never lead him/her to reduce lifetime utility in every state of the world. Since dynamic quantile preferences are monotonic, they are immune to this criticism.

\end{appendices}

\end{document}